\documentclass[lettersize,journal]{IEEEtran}
\usepackage{amsmath,amsfonts}
\usepackage{algorithm}
\usepackage{algpseudocode}
\usepackage{array}
\usepackage[caption=false,font=normalsize,labelfont=sf,textfont=sf]{subfig}
\usepackage{textcomp}
\usepackage{stfloats}
\usepackage{url}
\usepackage{verbatim}
\usepackage{graphicx}
\usepackage{enumitem}
\usepackage{eurosym}
\usepackage{tabularray}

\usepackage[caption=false,font=normalsize,labelfont=sf,textfont=sf]{subfig}
\usepackage{hyperref}
\hypersetup{
    colorlinks = true,
    citecolor = blue,
    linkcolor = blue,
    urlcolor = blue,
    bookmarksopen = true,
    filecolor=blue
}

\usepackage{booktabs}
\usepackage{multirow}

\newcommand{\soc}{\textit{SoC}}
\newcommand{\dep}{\textit{dep}}
\newcommand{\arr}{\textit{arr}}
\newcommand{\ch}{\textit{ch}}
\newcommand{\dis}{\textit{dis}}
\newcommand{\cs}{\textit{cs}}

\newcommand{\tr}{\textit{tr}}
\newcommand{\set}{\textit{set}}
\newcommand{\tot}{\textit{tot}}

\newcommand{\AC}{\textit{AC}}
\newcommand{\DC}{\textit{DC}}
\newcommand{\cyc}{\textit{cyc}}

\newcommand{\cale}{\textit{cal}}
\newcommand{\total}{\textit{tot}}

\newcommand\figref{Fig.~\ref}
\providecommand{\bosym}{\boldsymbol}

\def\BibTeX{{\rm B\kern-.05em{\sc i\kern-.025em b}\kern-.08em
    T\kern-.1667em\lower.7ex\hbox{E}\kern-.125emX}}

\usepackage{balance}
\begin{document}
\title{EV2Gym: A Flexible V2G Simulator for EV Smart Charging Research and Benchmarking}
\author{Stavros Orfanoudakis~\IEEEmembership{Student Member,~IEEE}, 
Cesar Diaz-Londono,~\IEEEmembership{Member,~IEEE}, 
Yunus E. Yılmaz,
Peter~Palensky,~\IEEEmembership{Senior Member,~IEEE}, Pedro~P.~Vergara,~\IEEEmembership{Senior Member,~IEEE}

\thanks{
This work used the Dutch national e-infrastructure with the support of the
SURF Cooperative using grant no. EINF-5716. Stavros is funded by the HORIZON Europe Drive2X Project 101056934.}
\thanks{
Stavros Orfanoudakis, Yunus E. Yılmaz, Peter Palensky, and Pedro P. Vergara are with the Intelligent Electrical Power Grids (IEPG) Section, Delft University of Technology, Delft, The Netherlands (emails: {s.orfanoudakis, p.palensky, p.p.vergarabarrios}@tudelft.nl, yemreyilmaz8@gmail.com).
}
\thanks{Cesar Diaz-Londono is with the Dipartimento di Elettronica, Informazione e Bioingegneria, Politecnico di Milano, Milan, Italy (email: cesar.diaz@polimi.it)}

}%

\markboth{}%
{Orfanoudakis S., et al. 2024}

\maketitle

\begin{abstract}
As electric vehicle (EV) numbers rise, concerns about the capacity of current charging and power grid infrastructure grow, necessitating the development of smart charging solutions. While many smart charging simulators have been developed in recent years, only a few support the development of Reinforcement Learning (RL) algorithms in the form of a Gym environment, and those that do usually lack depth in modeling Vehicle-to-Grid (V2G) scenarios. To address the aforementioned issues, this paper introduces the EV2Gym, a realistic simulator platform for the development and assessment of small and large-scale smart charging algorithms within a standardized platform. The proposed simulator is populated with comprehensive EV, charging station, power transformer, and EV behavior models validated using real data. EV2Gym has a highly customizable interface empowering users to choose from pre-designed case studies or craft their own customized scenarios to suit their specific requirements. Moreover, it incorporates a diverse array of RL, mathematical programming, and heuristic algorithms to speed up the development and benchmarking of new solutions.
By offering a unified and standardized platform, EV2Gym aims to provide researchers and practitioners with a robust environment for advancing and assessing smart charging algorithms.
\end{abstract}


\section{Introduction}
\label{sec:sec1}

\IEEEPARstart{T}{he} increasing number of electric vehicles (EVs) raises concerns about whether the current charging and power grid infrastructure can adequately accommodate them~\cite{Pandi2023ElectricVC}. Therefore, it is essential to evaluate the readiness of the existing infrastructure and develop smart charging solutions using accurate simulator platforms. Various problem formulations related to EV smart charging exist, each with its own unique characteristics and constraints~\cite{Das2020ElectricReview}. A variety of algorithms have been employed to tackle these challenges. Assessing the strengths and weaknesses of these solutions within a standardized simulator prior to real-world implementation can accelerate the integration of these methods into the practical operational management of EVs.

Classic mathematical optimization~\cite{Sengor2019OptimalUncertainty} algorithms and Model Predictive Control (MPC)~\cite{DIAZLONDONO2024100326} are effective for solving less complex EV optimization problems. In particular, mixed integer programming (MIP) has gained widespread attention in the optimization of EV charging scheduling, as it can model the complexity of the problem, e.g., EVs arriving and departing, while taking into account various constraints~\cite{Koufakis2020OfflineTransfer}, such as EV battery capacity, charging demand, and energy pricing~\cite{Meenakumar2020OptimalCapabilities}. 
However, mathematical programming encounters difficulties as the number of decision variables and constraints increases~\cite{Yang2023ASystems}, for example, when Charge Point Operators (CPOs) need frequent rerunning of optimization algorithms. This operational demand, especially with the anticipated increase of EVs in the near future, poses a significant efficiency challenge for mathematical programming approaches when confronted with complex, large-scale optimization problems.

Reinforcement Learning (RL) methods~\cite{SuttonReinforcementIntroduction} hold the potential to bridge the gap between optimality, linear modeling, uncertainty quantification, and scalability. In recent years, there has been a surge of interest in utilizing RL approaches to determine optimal EV charging behavior~\cite{QIU2023113052}. RL methods are capable of efficiently solving even the most complex problems, albeit sometimes at the expense of finding the optimal solution~\cite{9904958}. Therefore, the availability of standardized RL Gym environments for EV smart charging is crucial. Such environments can seamlessly support benchmarking for any type of open-source smart charging algorithm and facilitate the development of new specialized RL algorithms~\cite{brockman2016openai}.







\subsection{Existing EV Charging Simulators}
\label{sec:sec2}

\begin{table*}[t]
\caption{Overview of existing EV simulators focusing on smart charging strategies.}
\label{tab:simulators}
\resizebox{\textwidth}{!}{%
\begin{tabular}{@{}cccccccccc@{}}
\toprule
\begin{tabular}[c]{@{}c@{}}Simulator \\ Name\end{tabular} & V2G & \begin{tabular}[c]{@{}c@{}}Power Network\\  Impact\end{tabular} & \begin{tabular}[c]{@{}c@{}}EV \\ Models\end{tabular} & \begin{tabular}[c]{@{}c@{}}EV\\ Behavior\end{tabular} & \begin{tabular}[c]{@{}c@{}}Charging\\ Stations\end{tabular} & \begin{tabular}[c]{@{}c@{}}Available \\ Baseline Algorithms\end{tabular} & \begin{tabular}[c]{@{}c@{}}RL\\Ready\end{tabular} & \begin{tabular}[c]{@{}c@{}}Programming\\ Language\end{tabular} & Comments \\ \midrule
V2G-Sim~\cite{osti_1437011} & Yes & Partial & Diverse & \begin{tabular}[c]{@{}c@{}}Real Charging Transaction \\ Probability Distributions\end{tabular} & Uniform & \begin{tabular}[c]{@{}c@{}}- Heuristics, Mathematical\\  Programming\end{tabular} & No & \begin{tabular}[c]{@{}c@{}}Not\\ Open-source\end{tabular} & \begin{tabular}[c]{@{}c@{}}- Customizable \\ V2G simulations.\end{tabular} \\ \cmidrule(l){10-10} 
EVLibSim~\cite{rigas_evlibsim_2018} & Yes & No & Diverse & Randomized & Uniform & \begin{tabular}[c]{@{}c@{}}- Heuristics, Mathematical \\ Programming\end{tabular} & No & Java & \begin{tabular}[c]{@{}c@{}}- Easily customizable simulations\\  with a visual interface.\end{tabular} \\ \cmidrule(l){10-10} 
EV-EcoSim~\cite{balogun_ev-ecosim_2023} & Yes & Complete & Uniform & Randomized & Uniform & - Mathematical Programming & No & Python & - Grid-Impact analysis \\ \cmidrule(l){10-10} 
evsim~\cite{evsim} & No & Partial & Diverse & \begin{tabular}[c]{@{}c@{}}Real Charging Transaction \\ Probability Distributions\end{tabular} & Uniform & \begin{tabular}[c]{@{}c@{}}- Heuristics, Mathematical \\ Programming\end{tabular} & No & R & \begin{tabular}[c]{@{}c@{}}- Simulate and analyze the\\  charging behavior of EV users.\end{tabular} \\ \cmidrule(l){10-10} 
OPEN~\cite{MORSTYN2020115397} & Yes & Complete & Uniform & Randomized & Uniform &  \begin{tabular}[c]{@{}c@{}}- Heuristics,\\ Mathematical Programming\end{tabular} & No & Python & \begin{tabular}[c]{@{}c@{}}- Modelling, control \& simulations \\ for smart local energy systems.\end{tabular} \\ \cmidrule(l){10-10} 
ACN-Sim~\cite{lee_acn-sim_2021} & No & Complete & Uniform & \begin{tabular}[c]{@{}c@{}}Real Charging\\  Transactions\end{tabular} & Uniform & \begin{tabular}[c]{@{}c@{}}- Heuristics, MPC, RL, \\ Mathematical Programming\end{tabular} & Yes & Python & \begin{tabular}[c]{@{}c@{}}- Designing a complete\\ simulator framework.\end{tabular} \\ \cmidrule(l){10-10} 
SustainGym~\cite{yeh2023sustaingym} & No & Complete & Uniform & \begin{tabular}[c]{@{}c@{}}Real Charging\\  Transactions\end{tabular} & Uniform & - RL & Yes & Python & \begin{tabular}[c]{@{}c@{}}- Providing a benchmark for \\ sustainable RL applications.\end{tabular} \\ \cmidrule(l){10-10} 
Chargym~\cite{karatzinis_chargym_2022} & Yes & Very Limited & Uniform & Randomized & Uniform & - Heuristics, RL & Yes & Python & \begin{tabular}[c]{@{}c@{}}- Comparing RL algorithms \\ for smart charging.\end{tabular} \\ \cmidrule(l){10-10} 
\textbf{EV2Gym (Ours)} & Yes & Partial & Diverse & \begin{tabular}[c]{@{}c@{}}Real Charging Transaction \\ Probability Distributions\end{tabular} & Diverse & \begin{tabular}[c]{@{}c@{}}- Heuristics, MPC, RL,\\ Mathematical Programming\end{tabular} & Yes & Python & \begin{tabular}[c]{@{}c@{}}- Comprehensive simulator \\ for any control algorithm.\end{tabular} \\ \bottomrule
\end{tabular}%
}
\end{table*}

The importance of developing and evaluating algorithmic solutions that can support the influx of hundreds of thousands of EVs in the coming years has led to the development of many EV simulator platforms. However, existing platforms are either outdated, w.r.t. the simulation models implemented (V2G, battery degradation, etc.) and the simulator, or have limited simulation capabilities. Table~\ref{tab:simulators} provides a comprehensive comparison of existing EV simulator platforms, highlighting the advantages and the limitations of each one. The main comparison points revolve around the models developed, the EV behavior data, the study of the impact and the constraints of the power network, the developed algorithms, and their suitability for the development of RL algorithms.

V2G-Sim~\cite{osti_1437011} is a traditional EV simulator with rich modeling capabilities, such as EV models, EV behavior, etc. Still, the simulator is not open-source and does not provide an environment for developing RL algorithms. EVLibSim~\cite{rigas_evlibsim_2018} is another high-level optimal EV charging simulator for evaluating scenarios such as V2G, battery swapping, and inductive charging. This simulator focuses on providing diverse EV models but does not simulate the impact on the grid and by being written in Java it limits the accessibility to advanced machine learning packages developed in Python.
Furthermore, EV-EcoSim~\cite{balogun_ev-ecosim_2023} is an EV simulator environment focusing on the impact of charging on the distribution network. Additionally, it includes detailed battery utilization and degradation models but does not focus on realistic EV specification and behavior data.
EVsim~\cite{evsim} is a simulator built with realistic EV behavior data from public charging transactions focusing on assessing and analyzing the behavior of many EV users. Meanwhile, it does not include the option to research V2G and its impact on the grid.
Furthermore, OPEN~\cite{MORSTYN2020115397} is an integrated modeling, control, and simulation framework for smart local energy systems, including EVs, also solving power flow at a distribution system level. However, it only has uniform EV specification and behavior models and is not suitable for the development of RL algorithms.
Notably, these simulators lack the depth required for robust RL algorithm development due to either the absence of a standardized RL environment (Gym) or the oversimplified nature of their models.

To address this issue, most modern simulators also include a Gym environment that can seamlessly enable the testing of any RL algorithm.
ACN-Sim~\cite{lee_acn-sim_2021} is one of the most established EV simulator platforms, as it is developed following real EV parking lot characteristics while also including a Gym environment. It is also one of the only EV-specific simulators supporting complete power network calculations using open-source grid simulators, such as Pandapower and MatPower. The main limitation of ACN-Sim is that it was not designed with V2G support in mind; hence, it is not suitable for V2G research.
SustainGym~\cite{yeh2023sustaingym} is an extension of ACN-Sim that tries to standardize the EV optimal charging problem as an RL benchmark suite by defining the state and action space; however, without adding substantial functionality to the simulation.
Chargym~\cite{karatzinis_chargym_2022} is another EV simulation environment that can be used to develop RL algorithms focused on cost and penalty design; however, the underlying models are very simplified. For example, all EVs and chargers have identical specifications, and the EV behavior is not based on real data. Overall, there are Gym environments that provide high-quality EV charging management simulation; however, each one has different areas of focus or is outdated. Therefore, it is important to develop a standardized platform that can support the development of any type of control algorithm and that is also capable of simulating V2G scenarios.

\subsection{Our Contributions}
To address these gaps, we introduce EV2Gym\footnote{Access the open-source code, along with baseline algorithms and custom case studies, at \url{https://github.com/StavrosOrf/EV2Gym} and \url{https://github.com/distributionnetworksTUDelft/EV2Gym} }. This innovative V2G simulator is tailored for developing and evaluating smart charging algorithms, including rule-based, mathematical programming, metaheuristic, and RL. Unlike existing tools, EV2Gym incorporates realistic CPO assumptions, a critical aspect often overlooked in other approaches. To enhance the fidelity of our simulator, we populate it with highly detailed models of EVs, charging stations, and EV behaviors using real open-source data.
By offering a unified and standardized platform, EV2Gym aims to provide researchers and practitioners with a robust environment for the advancement and assessment of charging algorithms.
The contributions of this paper can be summarized as:
\begin{itemize}
    \item Flexible simulation environment for benchmarking and impact assessment of any charging algorithms.
    \item Detailed modeling of V2G EV charging management problems such as power setpoint tracking, profit maximization, utilization of PV, and demand response events.
    \item Fully customizable and easily configurable simulations allowing for highly detailed specifications of simulations with respect to charging topology and specifications, EV characteristics and behavior, prices, loads and generation on the level of the power transformer, etc.
\end{itemize}

\section{The EV2Gym Simulator}
\label{sec:sec3}

\begin{figure}[!t]
    \centering
    \includegraphics[width=0.8\linewidth]{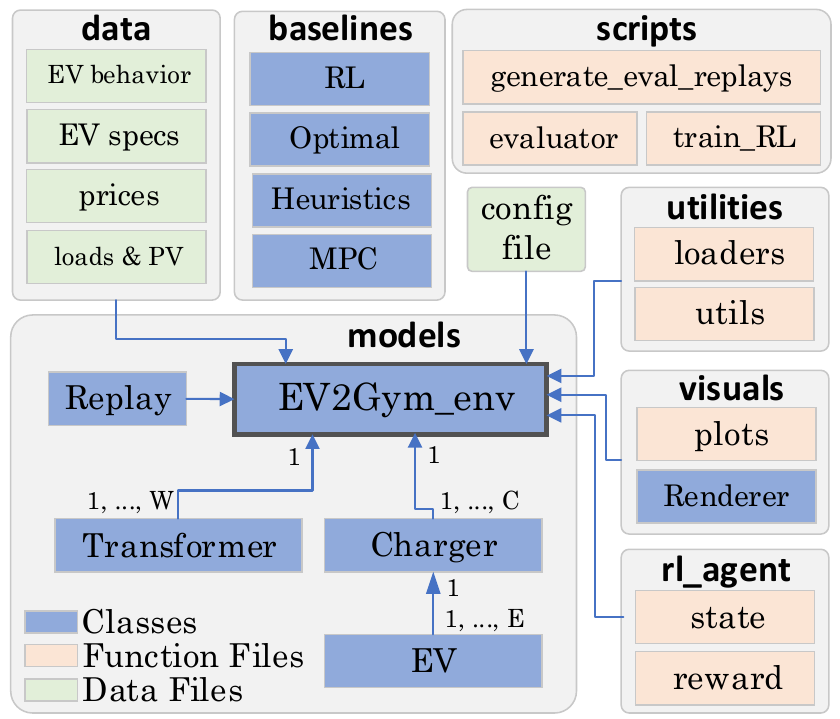}
    \vspace{-3mm}
    \caption{Directory and file structure of the EV2Gym package.}
    \label{fig:class_diagram}
\end{figure}

EV2Gym is an open-source simulator environment for comprehensive V2G simulations focused on developing and evaluating charging strategies to assess their performance and limitations. Furthermore, EV2Gym integrates with the Gym API~\cite{brockman2016openai}, streamlining the assessment of RL algorithms.
The directory and file structure of the EV2Gym package is illustrated in Fig.~\ref{fig:class_diagram}, revealing the interconnection of its underlying classes, functions, and data files.
The core component of EV2Gym is the \verb|EV2Gym_env| class, which contains lists of associated entities such as \verb|Transformers|, \verb|Chargers|, and \verb|EVs|. Additional essential functionalities are encapsulated within the \verb|utilities|, \verb|visuals|, and \verb|rl_agent| directories. The \verb|data| folder contains all the static data required for the simulation process.
Furthermore, the \verb|baselines| folder encompasses implementations of various smart charging algorithms, while the \verb|scripts| directory holds utility functions for efficiently evaluating any case study.

Most importantly, EV2Gym is modular; thus allowing for the seamless addition of extra functionalities.
This modularity is crucial, as open-source software empowers researchers to effortlessly extend the simulator to meet their specific requirements.
This section introduces the components of EV2Gym and provides a detailed description of the simulation flow and its underlying models, i.e., EV, Charger, and Transformer.

\begin{figure*}
    \centering
    \includegraphics[width=1\textwidth]{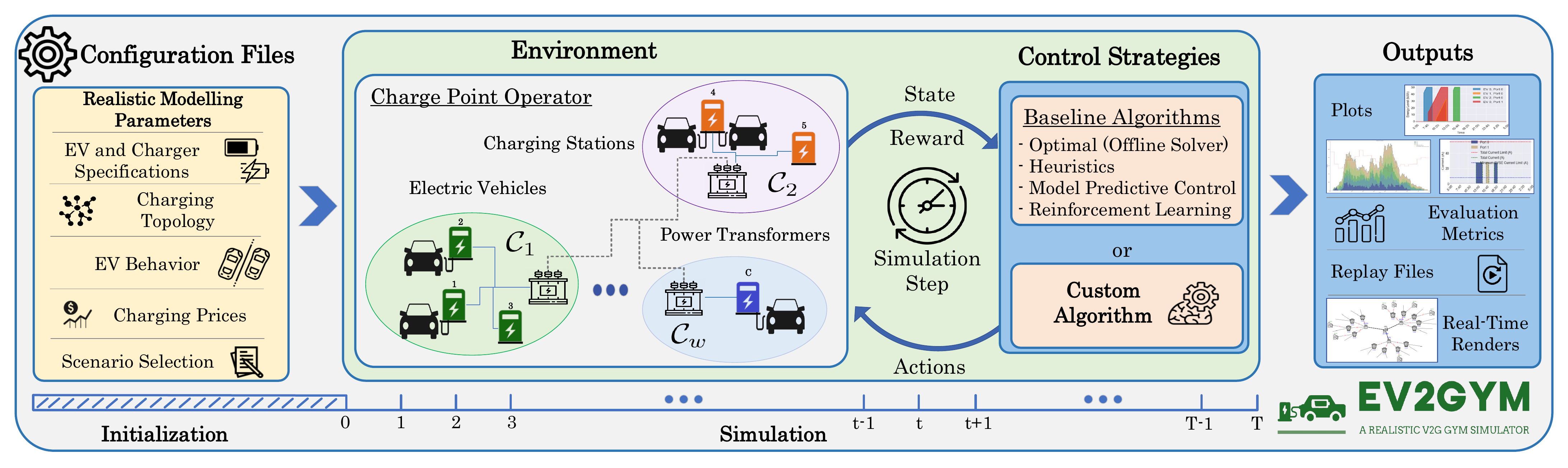}
    \vspace{-7mm}
    \caption{An EV2Gym simulation comprises three phases: the configuration phase, which initializes the models; the simulation phase, which spans $T$ steps, during which the state of models like EVs and charging stations is updated according to the decision-making algorithm; and finally, in the last phase, the simulator generates evaluation metrics for comparisons, produces replay files for reproducibility, and generates real-time renders for evaluation. }
    \label{fig:arch}
\end{figure*}

\subsection{Simulation Flow \& Configuration}

Executing a simulation is straightforward, requiring just a few lines of code, as depicted in Alg.\ref{alg:ev_simulation}. As a result, the simulation can be broken down into three phases, depicted in \figref{fig:arch}: Initialization, Simulation, and Evaluation.
The first phase encompasses the initialization of simulator models, empowering users to modify simulation parameters (e.g., \verb|EV2Gym(config_file)|), such as the simulation length $T$, timescale $\Delta t$, and charging topology configuration.
Furthermore, the user can fully customize the parameters of individual simulation models, including EVs, chargers, and transformers. Additionally, users can integrate custom time-series data, e.g., electricity prices and inflexible load curves, enriching the simulation environment with real-world dynamics.
Table~\ref{tab:ev_specs} shows all the configuration parameters contained in a \verb|config_file|. 
\begin{algorithm}[t]
\caption{Python code for running a simulation.}
\label{alg:ev_simulation}
\begin{algorithmic}[1]
\State \textbf{from} EV2Gym.EV2Gym\_env \textbf{import} EV2Gym
\State env$\gets$ EV2Gym(config\_file) \Comment{Initialization}
\State state, \_ $\gets$ env.reset()
\State agent $\gets$ Algorithm(...) \Comment{User-defined algorithm}
\For{$t = 1$ to $T$} \Comment{Simulation Phase}
    \State actions $\gets$ agent.get\_action(env, state)
    \State state, reward, done, \_, stats $\gets$ env.step(actions)
\EndFor
\end{algorithmic}
\end{algorithm}

\begin{table}[!t]
\caption{Overview of the input parameters of the EV2Gym models, emphasizing the flexibility in the experimental design.}
\label{tab:ev_specs}
\resizebox{\columnwidth}{!}{%
\begin{tabular}{@{}clc@{}}
\toprule
Model & \multicolumn{1}{c}{Input Parameters} & Symbol \\ \midrule
\multirow{8}{*}{Simulation} & - Timescale (in minutes) & $\Delta t$ \\
 & - Simulation Length (in steps)& $T$ \\
 & - Starting Date and Time &  \\
 & - Charging Topology and Properties &  \\
 & - EV Properties &  \\
 & - EV Scenario (Residential, Work, or Public) & \\
 & - Set of EV Charging Stations & $\mathcal{C}$ \\
 & - Set of Power Transformers & $\mathcal{W}$ \\
  \midrule
\multirow{10}{*}{EV} & - Min. \& Max. AC Charging Power (kW) & $\underline{P}^{\ch}_{\AC}$,$\overline{P}^{\ch}_{\AC}$ \\
 & - Min. \& Max. DC Charging Power (kW) & $\underline{P}^{\ch}_{\DC}$,$\overline{P}^{\ch}_{\DC}$ \\
 & - Min. \& Max. Discharging Power (kW) & $\underline{P}^{\dis}$,$\overline{P}^{\dis}$ \\
 & - Min. \& Max. Battery Capacity (kWh) & $\underline{E},\overline{E}$ \\
 & - Charge \& Discharge Efficiency & $\eta^{\ch}, \eta^{\dis}$ \\
 & - Battery Capacity at Arrival (kWh) & $E^{\textit{arr}}$ \\
 & - Desired Capacity at Departure (kWh) & $E^{\textit{*}}$ \\
 & - CV/CC Transition SoC (\%) & $\tau$ \\
 & - Time of Arrival \& Time of Departure (t) & $t^{\arr}, {t}^{\dep}$ \\ \midrule
\multirow{7}{*}{\begin{tabular}[c]{@{}c@{}}Charging\\ Station\end{tabular}} & - Min. \& Max. Charging Station Current (A) & $\underline{I}^{\cs}, \overline{I}^{\cs}$ \\
& - Min. \& Max. EVSE Charging Current (A) & $\underline{I}^{\ch}, \overline{I}^{\ch}$ \\
 & - Min. \& Max. EVSE Discharging Current (A) & $\underline{I}^{\dis}, \overline{I}^{\dis}$ \\
 & - Voltage (V) \& Phases & $V$, $\phi$ \\
 & - Set of EVSEs & $\mathcal{J}$ \\
 & - Type of Charger (AC or DC) &  \\
 & - Charging \& Discharging Prices (\euro / kWh) & $c^\ch$, $c^\dis$ \\
 \midrule
\multicolumn{1}{l}{\multirow{5}{*}{Transformer}} & 
 - Min. \& Max. Power (kW) & $\underline{P}^{\tr}, \overline{P}^{\tr}$ \\
& - Inflexible Loads (kW) & $P^{\textit{L}}$ \\
& - Solar Power Generation (kW) & $P^{\textit{PV}}$ \\
& - Demand Response Event (kW) & $P^{\textit{DR}}$ \\
\multicolumn{1}{l}{} & - Set of Connected Charging Stations & $\mathcal{C}_w$ 
 \\ \bottomrule
\end{tabular}%
}
\end{table}

Once the configuration phase has finished the main simulation takes place.
The simulation is partitioned into discrete time steps represented by the set $\mathcal{T}$, where $t \in \mathcal{T}$. During a simulation, there is a fixed number of charging stations denoted by $\mathcal{C}$, where $i \in \mathcal{C}$, and is connected to a transformer $w \in \mathcal{W}$, while the number of the connected EVs changes dynamically based on the simulation time and the scenario. Additionally, each charging station has a set of Electric Vehicle Supply Equipment (EVSE) $\mathcal{J}$, where $j \in \mathcal{J}$, wherein each one an EV $k$ from the set of available EVs $\mathcal{E}$ can connect. The goal of a simulation is to assess the performance of various charging strategies. Therefore, the user's charging strategy controls the current $I_{j,i,t}$ flowing from a charging station to a connected EV. For example, in a simulation with a set of charging stations $\mathcal{C}$ and each with a set of EVSEs $\mathcal{J}$ the control actions at timestep $t$ is a vector $\bosym I_t = [I_{j,i}] $, $\forall j \in \mathcal{J}$ and $\forall i \in \mathcal{C}$. 

As depicted in Alg.~\ref{alg:ev_simulation}, a simulation step is split into taking an action based on a user-defined algorithm (\verb|agent|), and then updating the state of the environment's models (\verb|env.step|) conditioned on the action taken. In EV2Gym the control algorithm has complete access to the state of the environment, while it can also access future information, such as electricity prices and EV schedules (arrival, departure, and SoC). However, users can define which part of the input is observable based on the case study they are exploring. 
Most importantly, the EV2Gym is designed to support any type of algorithm, i.e., heuristics, RL, MPC, or mathematical programming.

Finally, the simulation ends after $T$ steps have passed. At that time, the evaluation metrics (\verb|stats| in line 6 of Alg.~\ref{alg:ev_simulation}) of the simulation (see Sec.~\ref{sec:eval_metrics}), along with several figures, are generated in order to assess the performance of a charging algorithm. Moreover, a \verb|Replay| is generated, storing information about the environment configuration parameters and the EV schedules, so that the same simulation can be evaluated with alternative smart-charging algorithms. 


\subsection{Electric Vehicles}

Realistic EV models are necessary for precisely evaluating the effectiveness of charging strategies and the potential of V2G technology in facilitating the energy transition. Consequently, the \verb|EV| class is implemented to facilitate the simulation of diverse EV-related case studies. 

\begin{figure*}[t]
     \hfill
     \subfloat[Charging Type-2 AC.]{
         \centering
         \includegraphics[width=0.32\textwidth]{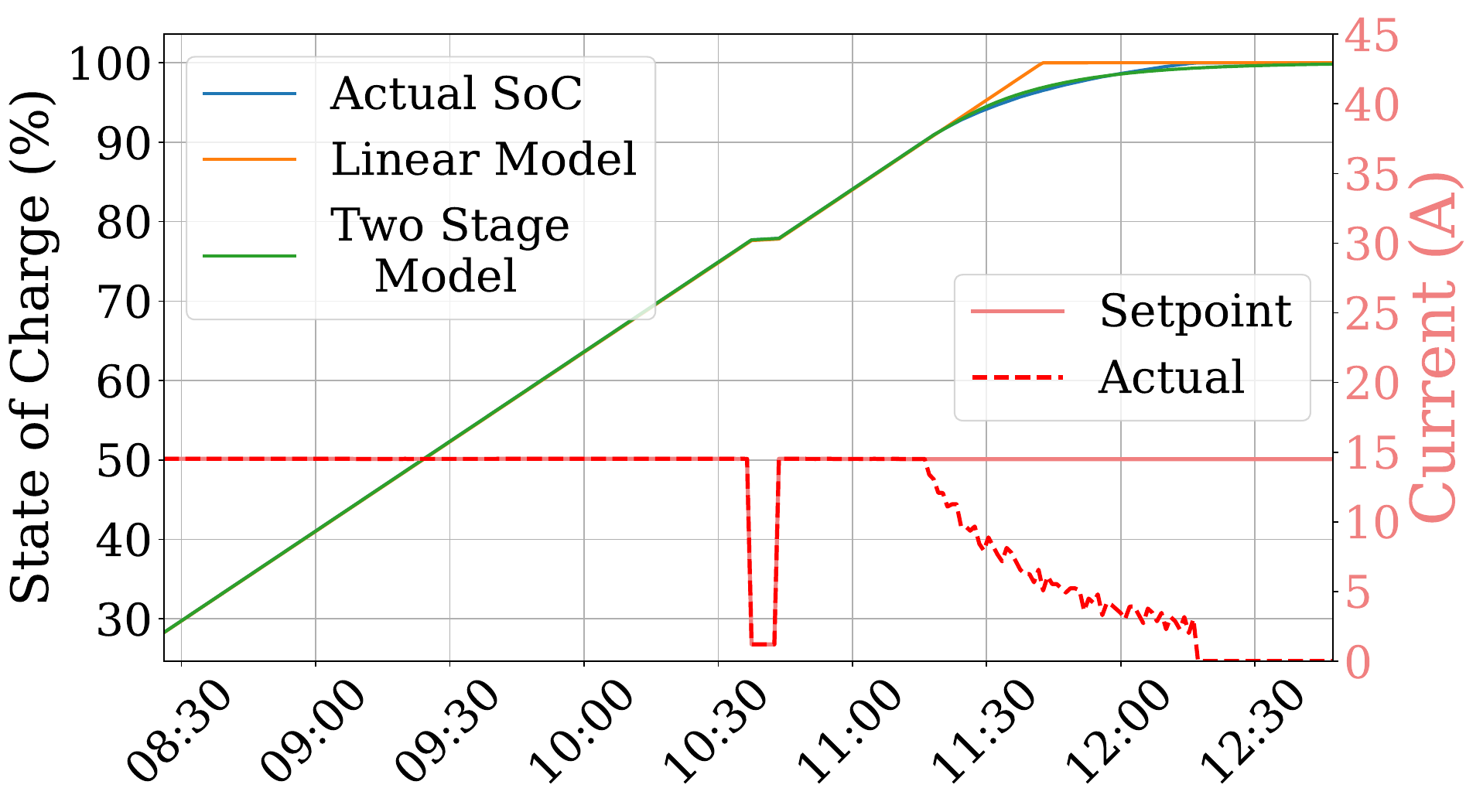}
         \label{fig:pf_a}
         }
     \subfloat[Charging DC 10 kW.]{
         \centering
         \includegraphics[width=0.32\textwidth]{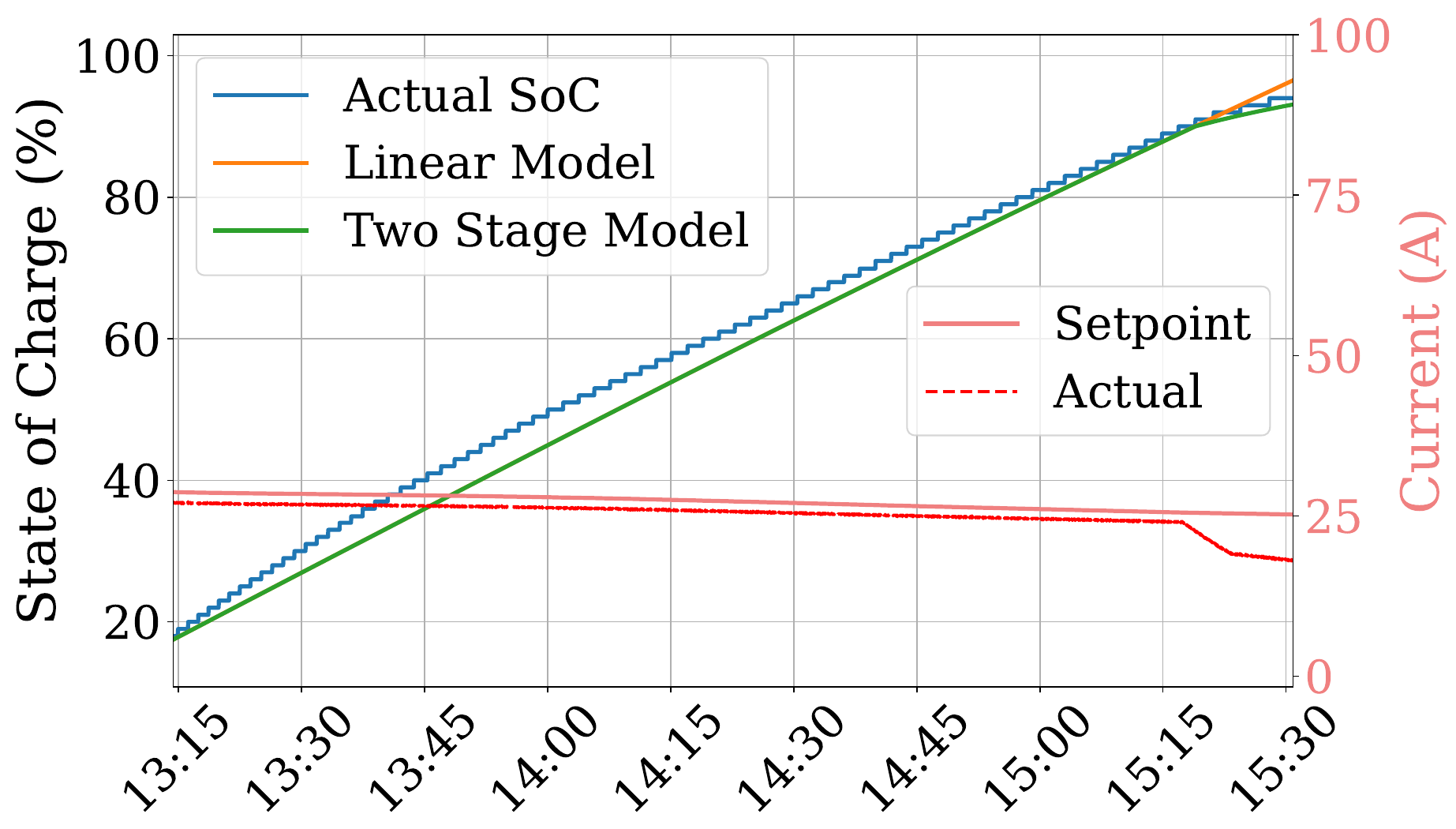}
         \label{fig:pf_b}}
    \subfloat[Discharging DC -10 kW.]{
         \centering
         \includegraphics[width=0.32\textwidth]{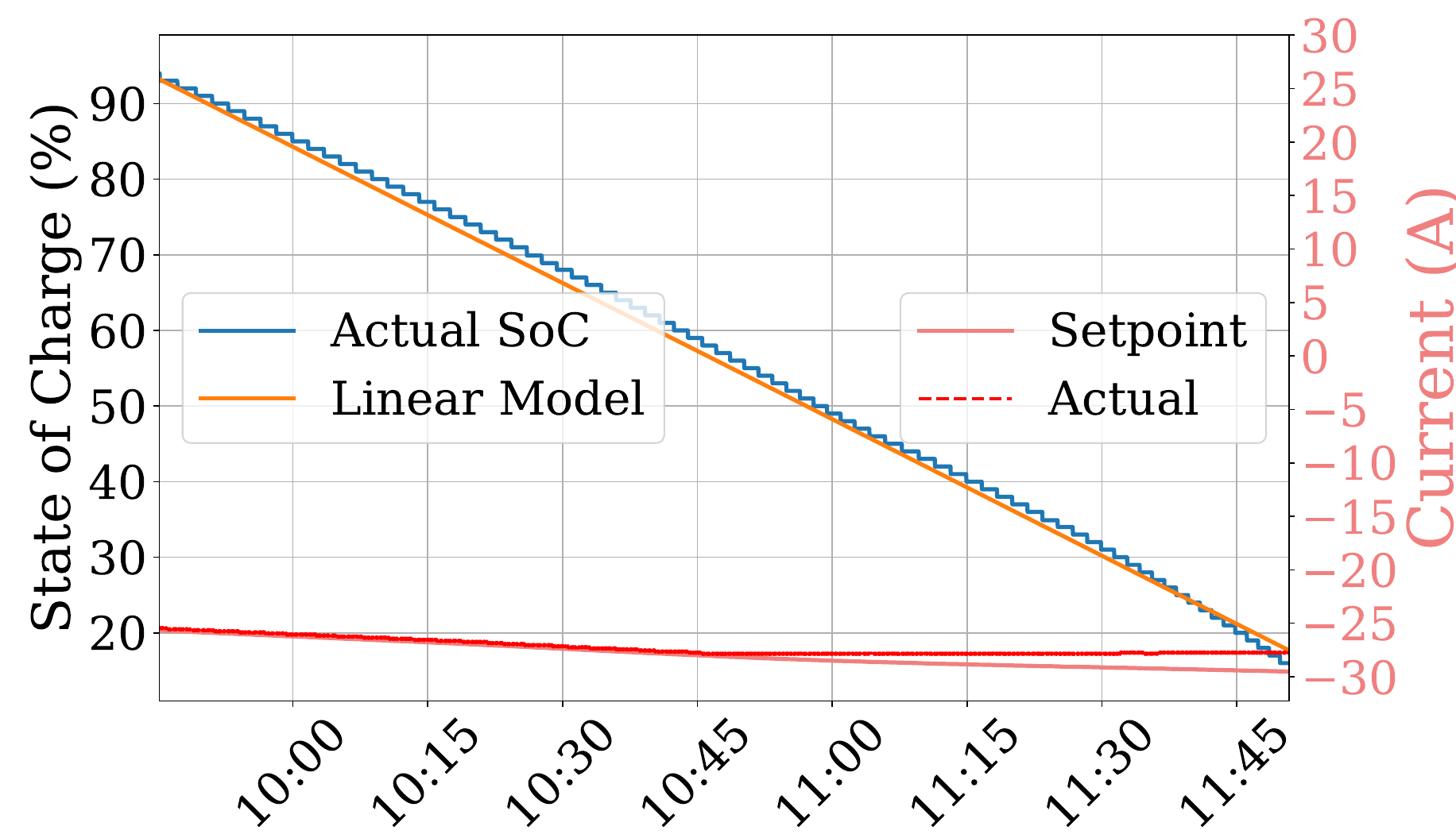}
         \label{fig:pf_c}}
         \hfill
         \vspace{-2mm}
        \caption{Exploring the relation between actual and modeled SoC as a function of current in AC and DC charging and discharging. 
        }
        \label{fig:charging curves}
\end{figure*}

\subsubsection{Charging and Discharging Models}
Similar to real cases, each EV can have different minimum and maximum charging and discharging power limits, depending on the charging mode (AC or DC), the power electronic limitations of the onboard battery management system (BMS), and the charger. In EV2Gym, the EV models have been designed with this property in mind. Thus, the min and max power limits are defined as $\underline{P}^{\text\ch}$ and $\overline{P}^{\ch}$, depending on the AC or DC charging mode. Similarly, the discharging power limits are defined as $\underline{P}^{\dis}$, and $\overline{P}^{\dis}$. Additionally, each EV has a maximum battery capacity, represented as $\overline{E}$, and a lower limit $\underline{E}$, which is used when discharging since the BMS of some EVs does not allow discharging under a certain threshold. Also, EVs have a desired $\soc^*$ at departure.

In the EV2Gym, a configurable two-stage model is available. The charging and discharging power is 
\begin{equation}
    P_t = \eta \cdot I_t \cdot V_t \cdot \sqrt{\phi},
\end{equation}
where $I_t$ is the control algorithm current, and $\eta$ refers to the charging ($\eta^\ch$) and discharging ($\eta^\dis$) efficiency. Also, $P_t$ depends on the charging station voltage $V_t$ and the phases $\phi$. $P_t$ is also subject to the lower and upper power limits of the EV and the charging station.
In detail, the two-stage model is:
\begin{equation}
 SoC_t \left\{
\begin{array}{ll}
      \soc_{t-1} + P_t \cdot \Delta t/ \; \overline{E} & \soc_{t-1} < \tau \\
       1 + (\soc_{t-1} - 1) \cdot \textit{exp} {\frac{P_t \cdot \Delta t}{ \overline{E} (\tau -1)}}  & \soc_{t-1} \geq \tau\\
\end{array} 
\right. 
\end{equation}
where $\tau \in (0,1)$, is the SoC transition threshold signifying the start of the constant voltage region~\cite{lee_acn-sim_2021}. Notice that if $\tau=1$, the model is linear. \figref{fig:charging curves} demonstrates the charging and discharging curves of the two models compared to real lab measurements from~\cite{drive2x_d81}. As observed in \figref{fig:charging curves}, even though the current setpoint is practically constant the actual perceived current decreases after some point. The two-stage model can more accurately mimic the actual charging curve of Type-2 charging (\figref{fig:pf_a}), while there is a slight deviation in the DC curves (\figref{fig:pf_b} and \figref{fig:pf_c}). These findings affirm the capability of EV2Gym models to accurately simulate realistic EV charging and discharging curves.

\subsubsection{Battery Degradation}

\begin{figure}[!t]
    \centering
    \includegraphics[width=0.9\linewidth]{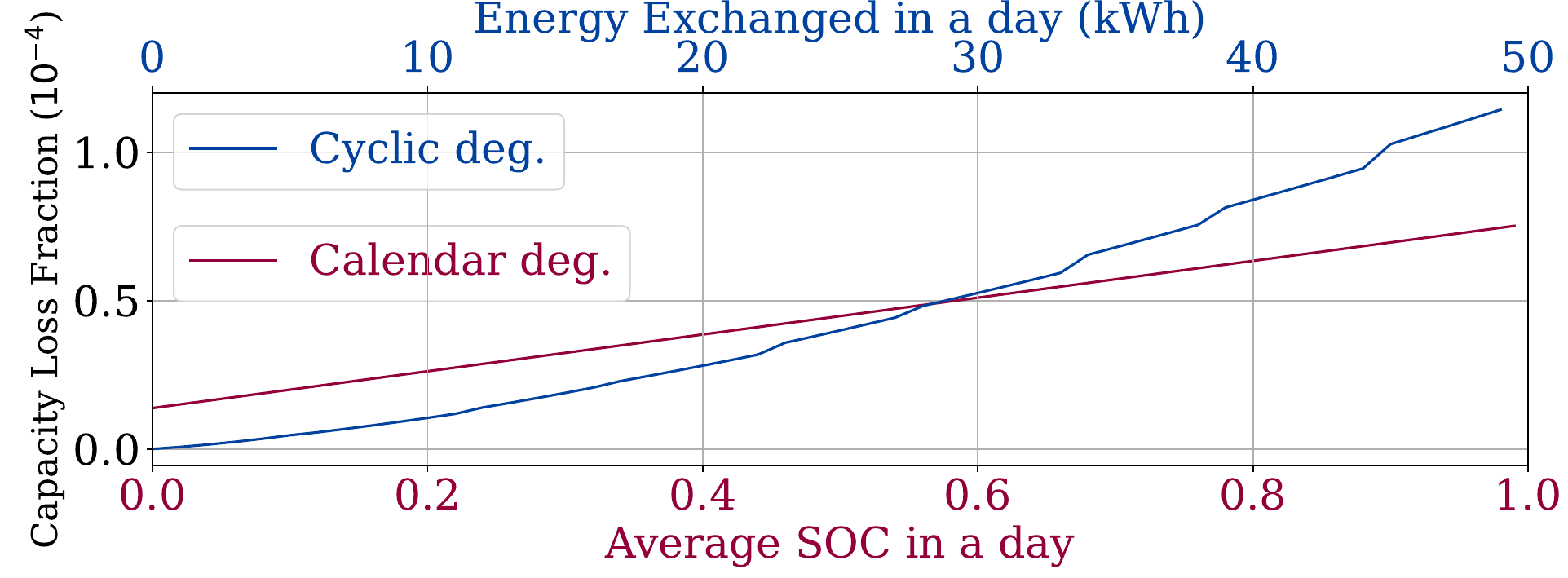}
    \vspace{-3mm}
    \caption{Visualizing calendar aging as a function of average SoC over a day, and cyclic degradation as a function of the total energy exchanged.}
    \label{fig:battery_degradation}
\end{figure}

V2G holds promise in supporting power network operations through its ability to offer flexible loads and provide users with financial incentives. Nevertheless, there are concerns regarding the impact of discharging on EV batteries. To address this, the EV2Gym incorporates a validated battery degradation model~\cite{LEE20236624}, enabling the evaluation of various charging strategies. The battery degradation model has a calendar ($d^{\cale}$) and a cyclic ($d^\cyc$) capacity loss component. Capacity loss due to calendar aging over a simulation period $T$ depends on the average $\langle{\soc}\rangle$ of the battery~\cite{SCHMALSTIEG2014325} and is defined as 
\begin{equation}
     d^\cale = 0.75 \cdot(\epsilon_0 \cdot\langle{\soc}\rangle - \epsilon_1)\cdot \text{exp} \left(-\frac{\epsilon_2}{\theta} \right) \cdot \frac{T}{(T^\total)^{0.25}},
\end{equation}
where $T^\total$ is the battery age in days, $\theta$ is the battery temperature, and $\epsilon_{0},\epsilon_{1},\epsilon_{2}$ are constants shown in Table~\ref{tab:constants}. The cyclic capacity loss depends on the total energy exchanged by the battery and the SoC at each simulation step
\begin{equation}
    d^\cyc = \left( \zeta_0 + \zeta_1 \frac{\int |\langle{\soc}\rangle- \soc(t)|\,\Delta t}{T}\right) \cdot \frac{\int |P(t)|\,\Delta t}{\sqrt{Q^\textit{acc}}}.
    \label{eq:dcyc}
\end{equation}
In~\eqref{eq:dcyc}, the numerator of the last fraction represents the battery throughput during the simulation, while $Q^\text{acc}$ represents the accumulated throughput during the battery's lifetime. The overall battery degradation is then defined as
\begin{equation}
    Q^\textit{lost} = d^\cale + d^\cyc.
\end{equation}
The fraction of the capacity loss over a single day for an EV with a 2-year-old, $50$ kWh battery is presented in \figref{fig:battery_degradation}. Battery degradation caused by calendar aging is highly dependent on the average SoC, while cyclic degradation can increase with the amount of energy exchanged.

\begin{table}[!t]
\centering
\caption{Battery Degradation Parameters and Values}
\label{tab:constants}
\resizebox{\columnwidth}{!}{%
\begin{tabular}{@{}cccccccc@{}}
\toprule
Param. & $\epsilon_0$ & $\epsilon_1$ & $\epsilon_2$ & $\theta$ & $z_0$ & $z_1$ & $Q^{acc}$ \\ \midrule
Value & $6.23\cdot10^6$ & $1.38\cdot10^6$ & $6976$ & $28$ & $4.02\cdot10^{-4}$ & $2.04\cdot10^{-3}$ & $11160$ \\ \bottomrule
\end{tabular}%
}
\end{table}

\subsubsection{EV Characteristics}
EV2Gym includes realistic EV specifications. In detail, the user has the option to define the specifications of the EV models (see Table~\ref{tab:ev_specs}) in a simulation either using realistic data from the RVO-NL~\cite{rvo2023sep} or custom ones.
In cases where users opt for realistic EV specifications from the RVO-NL report, the simulator dynamically samples from the available EVs listed in Table~\ref{tab:ev_specifications}, employing a probability distribution weighted by the total number of sales, each time a new EV is connected.

\begin{table}[t]
\centering
\caption{Total number of registered EVs in the Netherlands in $2023$}
\label{tab:ev_specifications}
\begin{tabular}{cccccc} 
\hline
EV Model        & \begin{tabular}[c]{@{}c@{}}Sales\\(2023 NL)\end{tabular} & \begin{tabular}[c]{@{}c@{}}$\overline{E}$\\~(kWh)\end{tabular} & \begin{tabular}[c]{@{}c@{}} $\overline{P}^{\ch}_{\AC}$ \\ (kW)\end{tabular} & \begin{tabular}[c]{@{}c@{}}$\overline{P}^{\ch}_{\DC}$ \\ (kW)\end{tabular} & \begin{tabular}[c]{@{}c@{}}$\overline{P}^{\dis}_{\DC}$ \\ (kW)\end{tabular}  \\ 
\hline
Tesla Model 3   & $45545$                                                     & $57.5$                                                            & $11$                                                                  & $170$                                                                 & -                                                                       \\
Kia Niro        & $23105$                                                     & $64.8 $                                                              & $11$                                                                  & $80$                                                                  & -                                                                       \\
Volkswagen ID.3 & $19950$                                                     & $58$                                                                 & $11$                                                                  & $120$                                                                 & -                                                                       \\
Hyundai Kona    & $17752$                                                     & $64$                                                                 & $11$                                                                  & $77$                                                                  & -                                                                       \\
Tesla Model Y   & $16186$                                                     & $57.5$                                                               & $11$                                                                  & $170$                                                                 & -                                                                       \\
Skoda Enyaq     & $16165$                                                     & $58$                                                                 & $11$                                                                  & $124$                                                                 & -                                                                       \\
Peugeot 208     & $14017$                                                     & $46.3$                                                               & $7.4$                                                                 & $101$                                                                 & -                                                                       \\
Renault Zoe     & $14008$                                                     & $52$                                                                 & $22$                                                                  & $46$                                                                  & -                                                                       \\
Volkswagen ID.4 & $13283$                                                     & $77$                                                                 & $11$                                                                  & $135$                                                                 & $10$                                                                      \\
Volvo XC40      & $12520$                                                     & $66$                                                                 & $11$                                                                  & $135$                                                                 & -                                                                       \\
Nissan Leaf     & $11977$                                                     & $39$                                                                 & $3.6$                                                                 & $46$                                                                  & $7$                                                                       \\
Tesla Model S   & $10899$                                                     & $75$                                                                 & $11$                                                                  & $250$                                                                 & -                                                                       \\
\hline
\end{tabular}
\end{table}

\subsubsection{EV User Behavior Models}
\label{sec:ev_user_beh}
EV2Gym uses authentic EV behavior data to mimic various case studies. The simulation starts with no EVs connected, gradually introducing them in each time step $t$ based on probability distributions for public, workplace, and residential load profiles provided by ElaadNL~\cite{elaad}. Users can specify the ElaadNL scenario to simulate, or can import custom EV behavior or charging transaction data from their private datasets. Specifically, whenever an EV is introduced at time $t^\arr$, the EV2Gym utilizes these distributions to determine both the departure time $t^\dep$ and the energy level upon arrival $E^\arr$, taking into account the time and day of arrival. 
\begin{figure*}[t]
    \centering
    \includegraphics[width=0.8\textwidth]{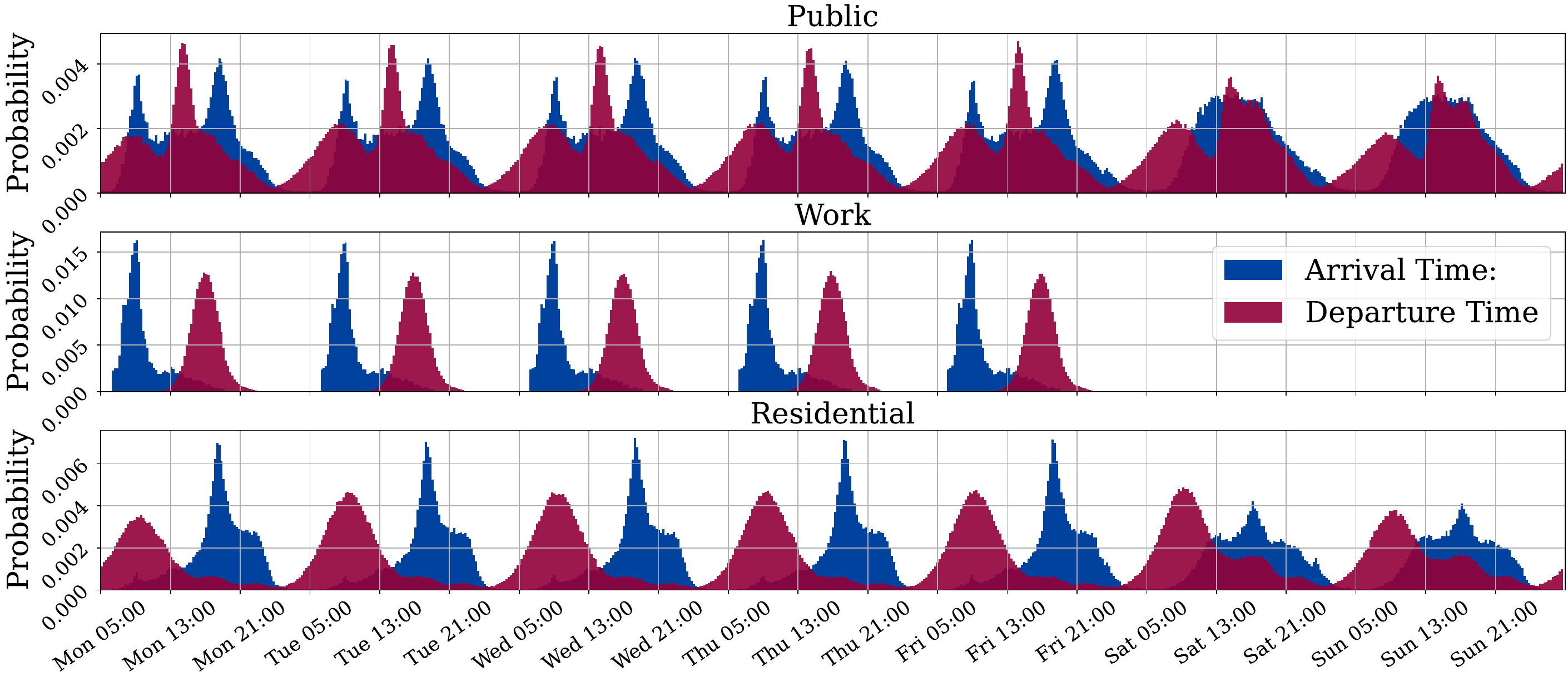}
    \vspace{-2mm}
    \caption{Distributions of arrival and departure times of EVs as a function of time and scenario (public, work, and residential).}
    \label{fig:arrival_and_departure}
\end{figure*}

\begin{figure}[t]
    \centering
    \includegraphics[width=1\linewidth]{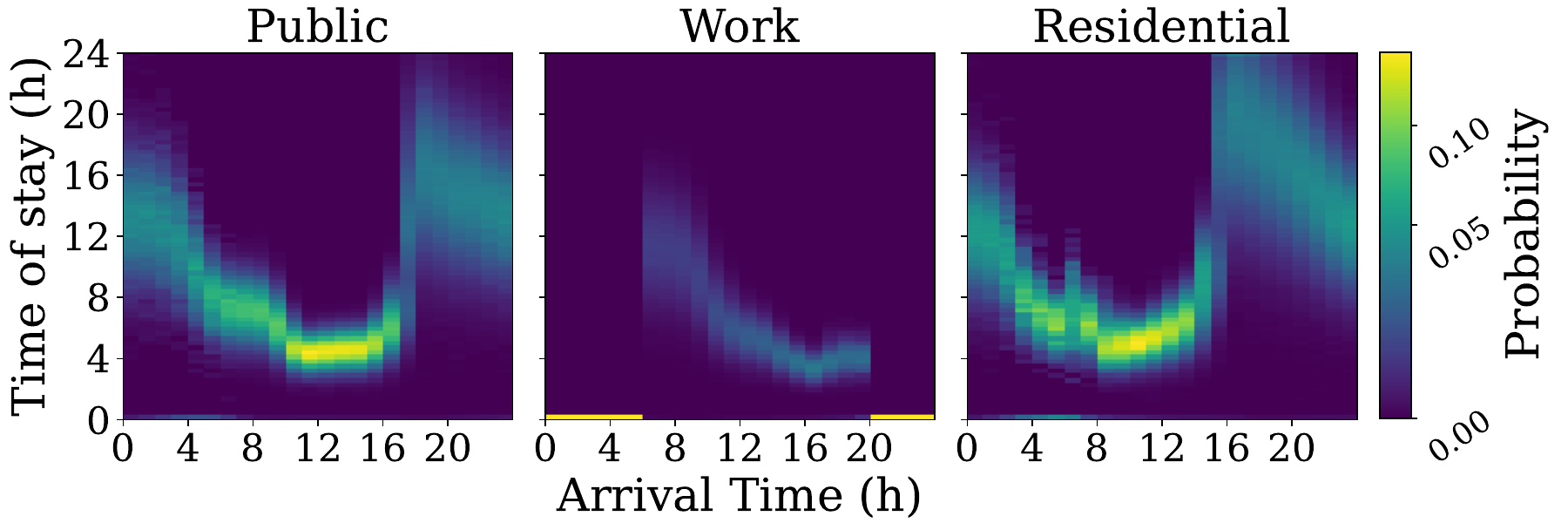}
    \caption{Probability density function of the duration of stay for each hour of arrival in a day. The brighter the color the higher the probability that an EV spawns with that value. Notice that, all values for a single hour sum up to 1.}
    \label{fig:time_of_stay}
\end{figure}

\begin{figure}[t]
    \centering
    \includegraphics[width=1\linewidth]{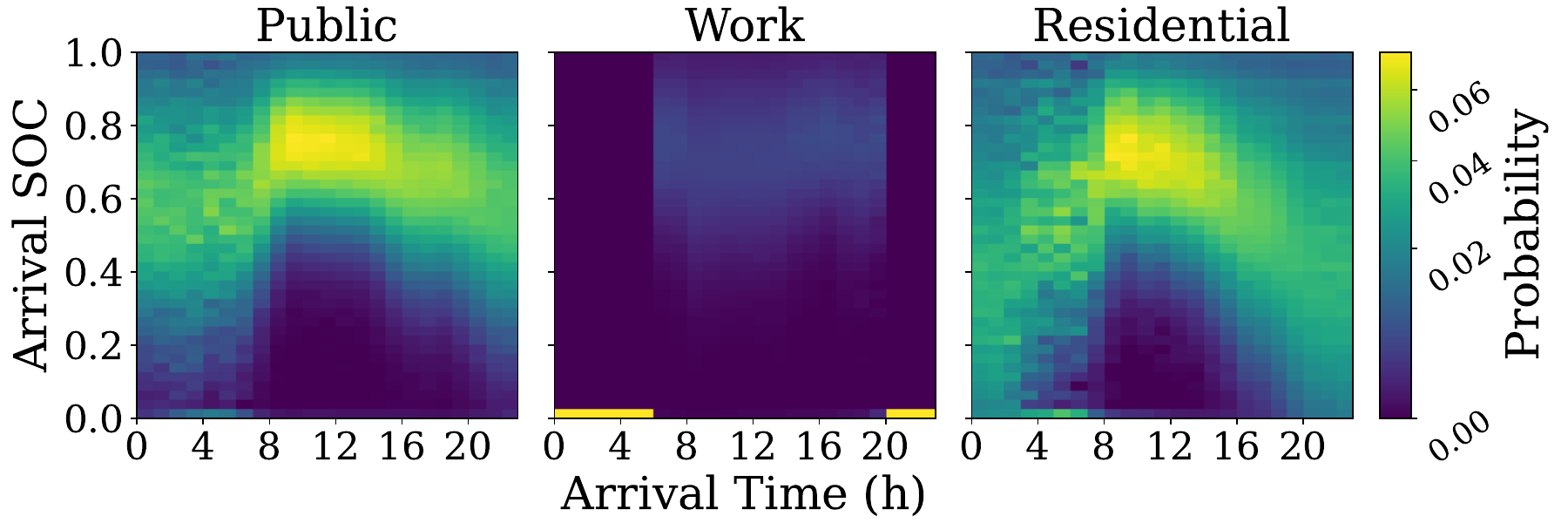}
    \caption{PDF of the SoC at arrival for each hour of arrival in a day.}
    \label{fig:arrival_soc}
\end{figure}

To validate the EV behavior model, we conducted extensive sampling, generating around 1 million charging transactions for each scenario.
The distribution of arrival and departure times over a week is presented in \figref{fig:arrival_and_departure}, showcasing variations in daily routines between weekdays and weekends. \figref{fig:time_of_stay} presents the probability density function (PDF) illustrating the duration of stay for each arrival time. Notably, in the workplace scenario, EVs do not arrive between 19:00-05:00, resulting in a zero stay time during these hours. Finally, \figref{fig:arrival_soc} showcases the PDF of SoC upon arrival for every hour.

\subsection{Charging Station}
In EV2Gym, each \verb|Charger| has a set of EVSE $\mathcal{J}$ where EVs can connect. Each charger can charge or discharge based on its voltage level $V$, number of phases $\phi^\cs$, and the total and EVSE current limitations (see Table~\ref{tab:ev_specs}). In practice, the total current of a charging station is limited as
\begin{equation}
    \underline{I}^{\cs}_{i} \leq 
    \sum_{j \in \mathcal{J}} I_{j,i} \leq
 \overline{I}^{\cs}_{i}.
 \label{eq:cs_current}
\end{equation}
Additionally, if the current control signal of an EVSE is between; $0$ and $\underline{I}^\ch$ (for charging), or 0 and $\underline{I}^\dis$ (for discharging), the actual current will be zero, because it is assumed that the EV's BMS will not allow for such low charging currents. Conversely, if the current signal exceeds the maximum EVSE current constraints, the BMS will limit the current to the maximum value. Also, if the sum of EVSE currents, $I_{j,i}$ for charger $i$, surpasses the charging station current limits, it is normalized down so that the operational constraint of \eqref{eq:cs_current} is not violated. Moreover, a charging station can have unique charging ($c^\ch$) and discharging ($c^\dis$) prices. EV2Gym uses by default the Dutch historic day-ahead energy prices, provided by entso-e~\cite{entsoe}, since they have a direct relation with the dynamic prices offered. If required, the user can include their own energy prices for their experiments.

\subsection{Power Transformer}

The current version of EV2Gym does not include power flow calculations; however, it aggregates sets of chargers $\mathcal{C}_w \subseteq \mathcal{C}$ at the level of power transformers $\forall w \in \mathcal{W}$. As shown in Table~\ref{tab:ev_specs}, a \verb|Transformer| has power limits denoted by $\underline{P}^{\tr}, \overline{P}^{\tr}$. A power transformer in EV2Gym has inflexible loads ($P^{\textit{L}}$) representing loads from houses or offices that cannot be shifted in time, solar power (PV) generation ($P^{\textit{PV}}$), and dynamic capacity reduction events ($P^{\textit{DR}}$), such as demand response events, that are communicated to the control algorithm only a few minutes ahead, e.g., 15 minutes ahead. The inflexible loads are based on randomized Pecan Street~\cite{pecanstreet} datasets and the PV generation profile is based on randomized Renewables.ninja~\cite{PFENNINGER20161251} data, while the occurrence of capacity reduction events is fully configurable by the user, with respect to start time, duration, etc. 
Therefore, the operational constraint for a power transformer $w$ is defined as 
\begin{equation}
\underline{P}^\tr_{w} \leq P^{\textit{EVs}}_{w} + P^{\textit{L}}_{w} + P^{\textit{PV}}_{w}\leq 
 \overline{P}^\tr_{w} - P^{\textit{DR}}_{w},
\label{eq:power_tr_constr}
\end{equation}
where $P^{\textit{EVs}}_{w}$ is the total power because of EVs. Notice that, unlike the charging station that normalizes the excess current, violations of operational constraints of transformers are feasible. Overloads are measured throughout the simulations as they are an important aspect of a smart charging algorithm. EV2Gym also generates forecasts of inflexible loads ($\widetilde{P}^{\textit{L}}$) and PV ($\widetilde{P}^{\textit{PV}}$) by sampling from a Gaussian distribution $\mathcal{N}(\mu,\sigma)$ with mean ($\mu$) the actual power consumed or generated at timestep $t$ and standard deviation ($\sigma$) defined by the user.





% ...




\section{Experimental Design}
\label{sec:sec4}

This section showcases two common EV smart charging problems, modified through the simulation configuration file (\verb|config_file|), to highlight the simulator's capabilities and inspire users to create their own case studies.

\subsection{Power Setpoint Tracking}
\label{sec:pst}

Scheduling and dispatching EVs often involves addressing challenges such as power setpoint tracking (PST), and capacity management. In practical scenarios, a CPO, or typically a company managing multiple EV chargers in their parking lot, either procures energy in advance from the day-ahead market or operates under a limited capacity contract. Consequently, they try to adhere closely to the procured power setpoint, ensuring efficient charging for all connecting EVs while fairly distributing energy among them. In this scenario, we assume that information about EV arrival and departure time, and SoC at arrival is unavailable.  However, it is assumed to be known when an EV is fully charged since the measured energy exchanged in a step is zero.

The PST problem has $T$ discrete timesteps represented by the set $\mathcal{T}$, where $t \in \mathcal{T}$. There is a fixed number of charging stations denoted by $\mathcal{C}$, where $i \in \mathcal{C}$, is connected to a transformer $w \in \mathcal{W}$, while the number of EVs changes dynamically. To model the presence of an EV, a binary variable $u_{j,i,t}$ is introduced, with $u_{j,i,t} = 1$, indicating that an EV is connected and ready to charge at EVSE $j$ at charging station $i$ during time $t$. Therefore, the mathematical formulation leads to a MIP problem that is described by \eqref{eq:opt3}-\eqref{eq:opt3.12}, $\forall  w \in \mathcal{W}, \; \forall  j \in \mathcal{J}, \; \forall  i \in \mathcal{C}, \;\forall t \in \mathcal{T}$. 
\begin{equation}
    \min_{I^\ch_{j,i,t},I^\dis_{j,i,t}} \sum_{t \in \mathcal{T}} \left(P^\set_t - P^\tot_t\right)^2
    \label{eq:opt3}
\end{equation}
Subject to:
\begin{flalign}
&P^\tot_{t} = \sum_{i \in \mathcal{C}} \sum_{j \in \mathcal{J}}{\left(P^\ch_{j,i,t} + P^\dis_{j,i,t}\right)}
  &\forall  j, \; \forall  i , \;\forall t&
    \label{eq:opt3.0}
\end{flalign}
\vspace{-5mm}
\begin{flalign}
& P^\ch_{j,i,t} = I^\ch_{j,i,t} \cdot V_{j,i,t} \cdot \sqrt{\phi_{j,i,t}} \cdot \eta^{\ch}_{j,i,t} \cdot \omega^\ch_{j,i,t}  & \forall  j, \; \forall  i , \;\forall t&
    \label{eq:opt3.1}
\end{flalign}
\vspace{-7mm}
\begin{flalign}
& P^\dis_{j,i,t} = I^\dis_{j,i,t} \cdot V_{j,i,t} \cdot \sqrt{\phi_{j,i,t}} \cdot \eta^{\dis}_{j,i,t} \cdot \omega^\dis_{j,i,t}  & \forall  j, \; \forall  i , \;\forall t&
    \label{eq:opt3.2}
\end{flalign}
\vspace{-7mm}
\begin{flalign}
& \underline{E}_{j,i} \leq E_{j,i,t} \leq \overline{E}_{j,i} & \forall  j, \; \forall  i , \;\forall t&
    \label{eq:opt3.3} 
\end{flalign}
\vspace{-7mm}
\begin{flalign}
& E_{j,i,t} = E_{j,i,t-1} + (P^\ch_{j,i,t} + P^\dis_{j,i,t}) \cdot \Delta t & \forall  j, \; \forall  i , \;\forall t&
\label{eq:opt3.4}
\end{flalign}
\vspace{-7mm}
\begin{flalign}
& E_{j,i,t} = E^{\textit{arr}}_{j,i,t} & \forall  j, \; \forall  i , \;\forall t | \; t = {t}^{\arr}_{j,i,t}&
\label{eq:opt3.5}
\end{flalign}
\vspace{-7mm}
\begin{flalign}
&\underline{I}^{\ch}_{j,i}\leq I^\ch_{j,i,t} \leq 
 \overline{I}^{\ch}_{j,i} & \forall  j, \; \forall  i , \;\forall t&
\label{eq:opt3.6}\\
&\underline{I}^{\dis}_{j,i} \geq I^\dis_{j,i,t} \geq 
 \overline{I}^{\dis}_{j,i} & \forall  j, \; \forall  i , \;\forall t&
\label{eq:opt3.7}
\end{flalign}
\vspace{-7mm}
\begin{flalign}
&I^\cs_{i,t} = \sum_{j \in \mathcal{J}}{\left(I^\ch_{j,i,t} \cdot \omega^\ch_{j,i,t} + I^\dis_{j,i,t} \cdot \omega^\dis_{j,i,t}\right)}
  & \forall  j, \; \forall  i , \;\forall t&
\label{eq:opt3.8}
\end{flalign}
\vspace{-5mm}
\begin{flalign}
&\underline{I}^{\cs}_{i} \leq I^\cs_{i,t} \leq 
 \overline{I}^{\cs}_{i} & \forall  i , \;\forall t&
\label{eq:opt3.9}
\end{flalign}
\vspace{-7mm}
\begin{flalign}
&  P^{\textit{EVs}}_{w,t} = \sum_{i \in \mathcal{C}_w} \sum_{j \in \mathcal{J}}{\left(P^\ch_{j,i,t} + P^\dis_{j,i,t}\right)} & \forall  w, \;\forall  j, \; \forall  i , \;\forall t&
\label{eq:opt3.101}
\end{flalign}
\vspace{-5mm}
\begin{flalign}
&\underline{P}^\tr_{w,t} \leq P^{\textit{EVs}}_{w,t} + P^{\textit{L}}_{w,t} + P^{\textit{PV}}_{w,t}\leq 
 \overline{P}^\tr_{w,t} - P^{\textit{DR}}_{w,t} & \forall  w, \; \forall t&
\label{eq:opt3.102}
\end{flalign}
\vspace{-7mm}
\begin{flalign}
& \omega^\ch_{j,i,t} + \omega^\dis_{j,i,t} \leq 1 & \forall  j, \; \forall  i , \;\forall t&
 \label{eq:opt3.11}
 \end{flalign}
\vspace{-6mm}
\begin{flalign}
& \omega^\ch_{j,i,t}=\omega^\dis_{j,i,t} = 0 & \forall  j, \; \forall  i , \;\forall t\left|u_{j,i,t}=0\right.&
    \label{eq:opt3.12}
\end{flalign}

This formulation aims to minimize the squared power tracking error \eqref{eq:opt3} by defining the charging and discharging current of the charging stations. Tracking error is the difference between the procured or setpoint power $P^\set_t$ and the actual power $P^\tot_t$ at time $t$. By minimizing this error, the costs of using unprocured energy and the losses of not using the procured energy are also minimized. The current of a single EVSE $j$ is defined by two different decision variables $I^\ch \cdot \omega^\ch,I^\dis \cdot \omega^\dis$ to model the discharging behavior differently than the charging behavior. The charging current and power ($I^\ch$ and $P^\ch$) are positive, while the discharging current and power ($I^\dis$ and $P^\dis$) take negative values. \eqref{eq:opt3.1} and \eqref{eq:opt3.2} refer to the power definitions and \eqref{eq:opt3.3}-\eqref{eq:opt3.5} refer to the EV battery constraints. \eqref{eq:opt3.6} and \eqref{eq:opt3.7} represent current charging and discharging constraints for each EV and EVSE, and \eqref{eq:opt3.9} for the whole charger. The transformer power constraint is described by \eqref{eq:opt3.102}. Finally, an EVSE cannot charge and discharge simultaneously; hence, the constraints in \eqref{eq:opt3.11}-\eqref{eq:opt3.12} for the binary variables $\omega^\ch$ and $\omega^\dis$.


\subsection{The V2G Profit Maximization Problem}
\label{sec:v2g_profit_max}

The second problem investigated focuses on maximizing a CPO's profits while ensuring that EV users' demands are fully satisfied.
In contrast to the PST problem, here, it is assumed that upon arrival at charging station $i$ and EVSE $j$, an EV shares its time of departure ($t_{j,i}^{\dep}$), and its desired battery capacity ($E_{j,i}^*$). Moreover, the battery capacity $E_{j,i,t}$ for each EV is known while it remains connected to a charger. These assumptions are usually considered in research as it is possible to retrieve this information from EVs, as more efficient communication protocols are introduced. The objective function is described in \eqref{eq:opt2} as a function of the charging ($c^\ch$) and discharging prices ($c^\dis$) for each EVSE $j,i$.
\begin{equation}
    \max_{I^\ch_{j,i,t},I^\dis_{j,i,t}} \sum_{t \in \mathcal{T}} \sum_{i \in \mathcal{C}} \left(-P^\ch_{i,t} \cdot c^\ch_{i,t} + P^\dis_{i,t} \cdot c^\dis_{i,t}\right) \cdot \Delta t
    \label{eq:opt2}
\end{equation}
Subject to constraints from \eqref{eq:opt3.1}-\eqref{eq:opt3.12} and:
\begin{flalign}
& E_{j,i,t} \geq E_{j,i,t}^{*} &\forall  j, \; \forall  i , \;\forall t \;| t = {t}^{\dep}_{j,i,t}&
\label{eq:opt2.1}
\end{flalign}
Moreover, in this scenario, the controller possesses knowledge of the inflexible load and PV forecasts to comply with the power transformer constraints \eqref{eq:opt3.102}.


\begin{table}[t]
\centering
\caption{A list of evaluation metrics provided by the simulator.}
\label{tab:eval_metrics}
\begin{tabular}{@{}ccc@{}}
\toprule
Symbol & Metric & Equation \\ \midrule
$E^\ch$ & \begin{tabular}[c]{@{}c@{}} -Total Energy \\ Charged (kWh)\end{tabular} & $\sum_{t \in \mathcal{T}}{P^{\ch}_t\cdot \Delta t}$\\
$E^\dis$ & \begin{tabular}[c]{@{}c@{}}- Total Energy \\ Discharged (kWh)\end{tabular} & $\sum_{t \in \mathcal{T}}{P^{\dis}_t\cdot \Delta t}$ \\
$\epsilon^\textit{usr}$ & \begin{tabular}[c]{@{}c@{}}- User \\ Satisfaction (\%)\end{tabular} & $\frac{1}{|\mathcal{E}|} \cdot \sum_{k \in \mathcal{E}}{ \frac{\soc_k}{\soc^*_k}}$ \\
c &\begin{tabular}[c]{@{}c@{}}- Total Profits \\\& Costs (\euro)\end{tabular} & $\sum_{t \in \mathcal{T}} \left(P^\ch_t \cdot c^\ch_t + P^\dis_t \cdot c^\dis_t\right) 
 \cdot \Delta t$
\\
 $\epsilon^\textit{ov}$ & \begin{tabular}[c]{@{}c@{}}- Transformer \\Overload (kWh)\end{tabular}  & $\sum_{t \in \mathcal{T}}\sum_{w \in \mathcal{W}}{\textit{max}(P^\tr_{w,t}- \overline{P}^\tr_{w,t},0) }$ \\
$\epsilon^{|\textit{tr}|}$ & \begin{tabular}[c]{@{}c@{}}- Tracking \\ Performance (kWh)\end{tabular}  & $\sum_{t \in \mathcal{T}}{ \left|P^{\set}_t - P^{\tot}_t\right| \cdot \Delta t}$ \\
$\epsilon^\textit{tr}$ & \begin{tabular}[c]{@{}c@{}} - Squared \\ Tracking Error\end{tabular} & $\sum_{t \in \mathcal{T}}{ \left(P^{\set}_t -P^{\tot}_t\right)^2}$ \\
$Q^\textit{lost}$ & \begin{tabular}[c]{@{}c@{}}- Total Battery \\ Capacity Loss\end{tabular} & $ \sum_{k \in \mathcal{E}} \left(d^{\cale}_k+ d^{\cyc}_k \right)$ \\ \bottomrule
\end{tabular}%
\end{table}
\subsection{Evaluation Metrics}
\label{sec:eval_metrics}

At the end of a simulation, the list of evaluation metrics, shown in Table~\ref{tab:eval_metrics} is generated (see \verb|stats| in Alg.~\ref{alg:ev_simulation}). Evaluation metrics can help in assessing the performance of various smart charging algorithms uncovering their strengths and weaknesses. Also, more evaluation metrics can be added to fit the needs of the users.




\section{Baseline Algorithms}

We offer a suite of baseline methods to expedite the comparison of smart-charging algorithms. These algorithms are categorized into heuristics, mathematical programming, and RL. This diverse selection allows for comprehensive evaluation across different algorithmic approaches.

\subsection{Heuristics}

Three simple rule-based algorithms are provided. The \textit{Charge As Fast As Possible} (AFAP) algorithm charges EVs with maximum power immediately after they connect without complying with transformer-level constraints. Similarly, the \textit{Charge As Late As Possible} (ALAP) heuristic starts charging EVs at maximum speed as late as possible to reach the desired SoC. Finally, \textit{Round Robin} (RR) charges EVs in turns only up to the power setpoint ($P^{\textit{set}}$), ensuring each EV gets a fair share of energy in a cyclical manner. 

\subsection{Mathematical Programming \& MPC}
Mathematical programming and Model Predictive Control (MPC) algorithms are also provided. In detail, when solving optimization problems, it is important to compare the outputs of the developed algorithms with optimal solutions. For this reason, we provide Gurobi models capable of solving optimally, assuming complete knowledge, the PST and V2G profit maximization problem, introduced in Sec.~\ref{sec:sec4}. Moreover, real-time MPC solutions for the profit maximization problem are developed~\cite{8511498}. MPC is a dynamic approach, continuously recalibrating strategies to steer towards optimal outcomes, tackling the inherent uncertainties.

\subsection{Deep Reinforcement Learning}
Optimal EV dispatch problems are sequential decision-making problems under uncertainty; thus, they can be formulated as Markov Decision Processes (MDP) with ($\mathcal{S},\mathcal{A},\mathcal{P},\mathcal{R}$).
The state-space $\mathcal{S}$ can include information about electricity prices, grid demand, user preferences, SoC, etc., and the actions $\mathcal{A}$ for every EV are related to the charging or discharging power during a time period $t$. The unknown state-transition probability is described by $\mathcal{P}$, while the reward $\mathcal{R}$ function can be designed to maximize any objective function related to EV optimization problems. Naturally, MDPs can be solved using RL~\cite{SuttonReinforcementIntroduction} to learn the optimal policy by interacting with the environment and updating the policy based on observed rewards and states. EV2Gym is a Gym environment that can run any RL algorithm. In detail, we are utilizing the Stable Baselines 3 python package, which includes an implementation for the most common RL algorithms~\cite{stable-baselines3}: A2C, ARS, SAC, DDPG, TD3, TQC, PPO, TRPO, etc.
Therefore, implementing custom algorithms, including designing new reward and state functions is simple to do. Furthermore, the simulator allows for the adjustment of the action space as needed, facilitating the implementation of discrete actions.

\subsubsection{The PST Problem} 
Solving an EV optimal charging problem requires creating specific state and reward functions. In the case of PST problem, the state space is a vector consisting of $3+3|N|$ variables, where $|N|$ is the total number of EVSE where an EV can connect. In detail, at timestep $t$
\begin{equation}
    S_t=[t,P^\set_t,P^\tot_{t-1}] \cup [d_{j,i}, E_{j,i,t} - E^{\arr}_{j,i}, t - t^\arr_{j,i}] \; \;\;\; \forall j, \forall i
\end{equation}
with $d_{j,i}$ being 1 if the EV is fully charged and $0.5$ if the EV still receives energy.
If no EV is connected at EVSE $j$ of charging station $i$, then the EV state parameters are replaced by zeros.
Notice that in this case study we assume that we do not have prior information about EV arrivals and that the SoC is unknown because of the communication protocol. However, what is known is the total energy charged since the time of arrival ($E_{j,i,t} - E^{\arr}_{j,i}$), and the time of stay up to step $t$.
The reward function is $r_t = -\left(P^\set_{t-1} - P^\tot_{t-1}\right)^2$. To ensure bounded actions in RL, the action vector $\mathbf{a}$ is constrained within the interval $[0, 1]^N$, where a value of $1$ indicates charging at maximum power and $0$ denotes no action. 

\subsubsection{The V2G Profit Maximization Problem} 
Contrary to the realistic PST problem, here we assume we know information about EV departure and SoC as long as the EV connects to the charger. The charging energy prices, forecasted inflexible loads, and PV generation for a horizon $h$, are also considered to be known. Therefore, the state at timestep $t$ is defined as:
\begin{multline}
    S_t= [t, P^\tot_{t-1}, c^\ch_{t:t+h}] \cup [\widetilde{P}^{\textit{L}}_{w,t:t+h} - \widetilde{P}^{\textit{PV}}_{w,t:t+h},P^{\textit{DR}}_{w,t:t+h} ]  \\ \cup [\soc_{j,i},t^\dep_{j,i} - t] \; \;\;\;\forall w, \forall j, \forall i.
\end{multline}
The state has $1 + h + 2h|W| + 2|N|$ variables, where $W$ is the total number of power transformers. This state space has many more variables than the PST problem, since the V2G problem is a more complex optimization task. The reward function $\forall w, \forall j, \forall i$ is:
\begin{equation}
    r_t = c_{t-1} - 100 \cdot \epsilon_{w,t-1}^\textit{ov} - 100 \cdot \textit{exp}\left( -10 \cdot \epsilon_{j,i,t^\dep}^\textit{usr} \right),
    \label{eq:reward_f}
\end{equation}
where $c_{t-1}$ are the costs, $\epsilon_{w,t-1}^\textit{ov}$ are the power transformer overloads, and $\epsilon_{j,i,t^\dep}^\textit{usr}$ is the user satisfaction score of EVs at departure of the previous step, as defined in Table~\ref{tab:eval_metrics}. The function in \eqref{eq:reward_f} rewards profits while it penalizes heavily overloads and unsatisfied customers. The reward function in \eqref{eq:reward_f} was obtained after practical experimentation with alternative formulations. Similar to the PST problem, the action vector $\mathbf{a}$ is constrained within the interval $[-1, 1]^N$, where a value of $-1$ indicates discharging with maximum power, $1$ means charging with maximum power and $0$ denotes no action. 

\section{Experimental Evaluation}
\label{sec:sec5}

This section presents an experimental evaluation of three case studies, each assessed using all available baseline algorithms and compared using the proposed evaluation metrics.

\subsection{PST Benchmark}

\begin{table}[t]
\centering
\caption{Baseline algorithms for the PST problem with 20 Chargers.}
\label{tab:pst_comp}
\resizebox{\columnwidth}{!}{%
\begin{tabular}{@{}cccccc@{}}
\toprule
Algorithm & \begin{tabular}[c]{@{}c@{}}Energy \\ Charged\\ (kWh)\end{tabular} & \begin{tabular}[c]{@{}c@{}}$\epsilon^{\textit{usr}}$ (\%)\end{tabular} & \begin{tabular}[c]{@{}c@{}}Tracking \\ Error \\ ($10^{3}$)\end{tabular} & \begin{tabular}[c]{@{}c@{}}Energy\\ Error\\ (kWh)\end{tabular} & \begin{tabular}[c]{@{}c@{}}Reward\\ $(10^3)$\end{tabular} \\ \midrule
AFAP & $288$ ±$61$ & $100$ ±$0$ & $35.8$ ±$13.9$ & $313$ ±$59$ & $-18.2$ ±$7.5$ \\
RR & $276$ ±$59$ & $99$ ±$1$ & $4.2$ ±$2.3$ & $57$ ±$16$ & $-2.0$ ±$0.8$ \\
A2C & $144$ ±$41$ & $84$ ±$4$ & $30.5$ ±$12.0$ & $295$ ±$57$ & $-30.0$ ±$12.1$ \\
ARS & $255$ ±$53$ & $96$ ±$2$ & $17.4$ ±$7.6$ & $213$ ±$46$ & $-15.8$ ±$7.3$ \\
SAC & $281$ ±$59$ & $99$ ±$1$ & $20.1$ ±$8.6$ & $231$ ±$48$ & $-13.8$ ±$6.2$ \\
TD3 & $147$ ±$43$ & $84$ ±$4$ & $29.9$ ±$11.8$ & $292$ ±$56$ & $-29.4$ ±$11.8$ \\
TQC & $252$ ±$54$ & $96$ ±$2$ & $20.8$ ±$8.8$ & $237$ ±$49$ & $-17.4$ ±$8.0$ \\
TRPO & $79$ ±$30$ & $77$ ±$4$ & $25.2$ ±$11.6$ & $264$ ±$60$ & $-25.2$ ±$11.6$ \\ \midrule
Optimal & $287$ ±$60$ & $100$ ±$0$ & $0.6$ ±$0.3$ & $46$ ±$14$ & $-0.6$ ±$0.3$ \\ \bottomrule
\end{tabular}%
}
\end{table}

The first case study investigated is the public PST problem (see Sec.~\ref{sec:pst}) with $20$ chargers, using $15$-minute timesteps for almost a day ($85$ steps). 
Furthermore, the ``Public'' EV behavior model (Sec.~\ref{sec:ev_user_beh}) is employed, while the V2G functionality is not available, to replicate the charging behavior observed in publicly accessible charging stations.
The primary difficulty of this case arises from the lack of prior information about EV arrivals and the absence of real-time SoC data. Table~\ref{tab:pst_comp} presents the average performance of various baselines after $100$ stochastic experimental runs. Notice that, only the results from applicable baselines are presented because the ALAP, and MPC algorithms require prior information of the EV departure, which is assumed to be unknown to the CPO in this case study. A key metric is the energy error, which indicates the actual deviation from the power setpoint. Additionally, metrics such as total energy charged, user satisfaction, tracking error, and reward are examined. The final row illustrates the optimal solution assuming complete knowledge, serving as an experimental upper bound on solution quality, although it does not represent a realistic scenario. 
Similarly, the AFAP heuristic algorithm serves as an experimental lower bound, illustrating the outcomes when no smart charging strategy is employed. Consequently, we anticipate that all other algorithmic solutions will yield results that are inferior to the optimal solution yet superior to those generated by AFAP.

\begin{figure}[t]
    \centering
    \includegraphics[width=1\linewidth]{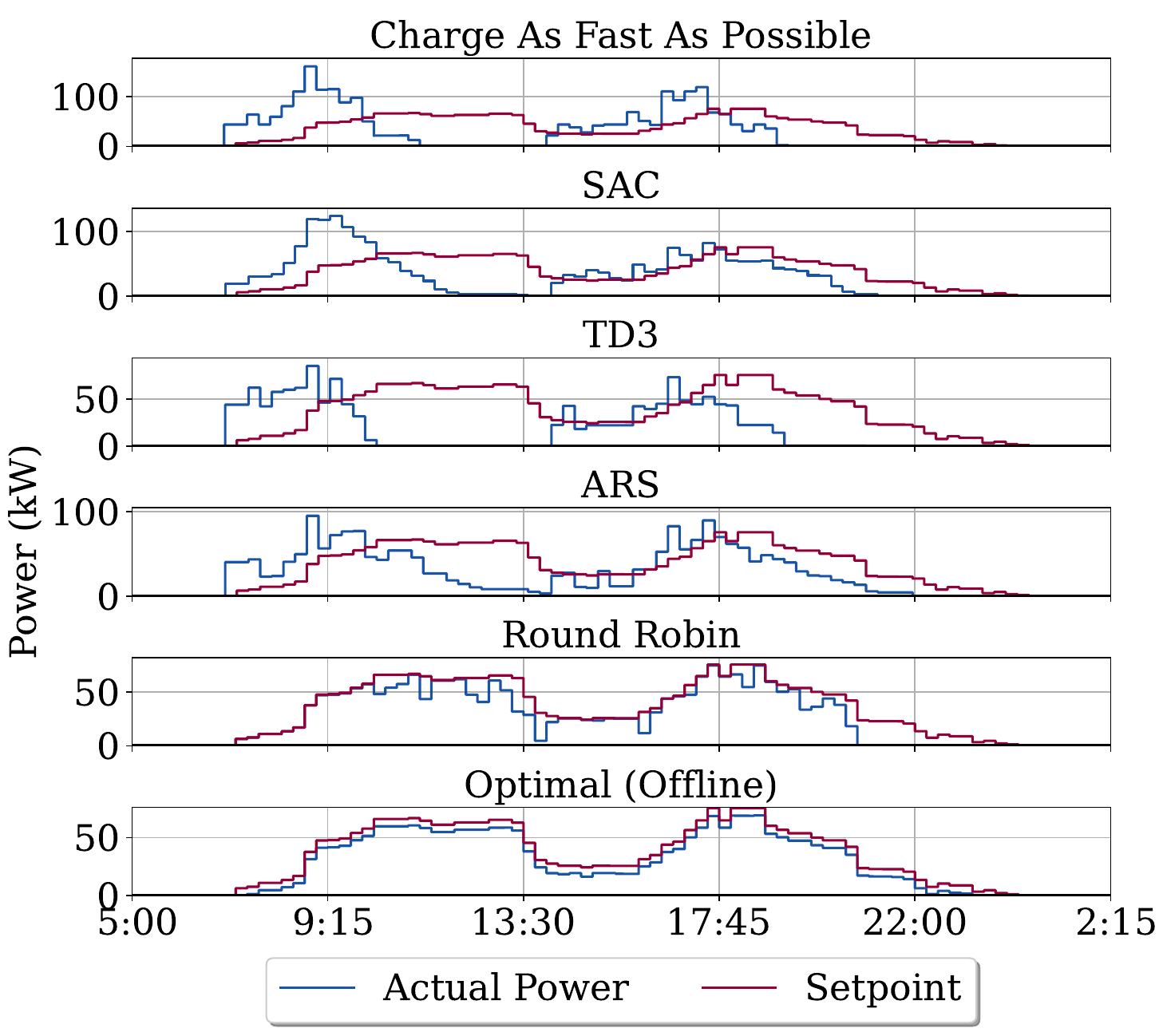}
    \vspace{-8mm}
    \caption{Side-by-side comparison of baseline algorithms for the PST problem for 20 EV chargers throughout a day (85 steps, 15 minutes per step).}
    \label{fig:pst_comp}
\end{figure}

As expected, AFAP has the worst average energy error ($313$ kWh) while RR has the best ($57$ kWh), excluding the optimal solution. The average energy error of RL baselines varies from $213$ to $295$ kWh, while the user satisfaction varies from $77\%$ to $99\%$. RR also maximizes the fair distribution of energy as shown by the $99\%$ user satisfaction. Even though RR seems to be the ultimate solution for the PST problem, it has limitations when the number of charging stations and operational constraints increase. \figref{fig:pst_comp} illustrates the performance of selected baseline algorithms for a single run.

\subsection{V2G Profit Maximization with Loads Benchmark}

The second case study is about maximizing the profits of V2G smart charging in a workplace parking lot while considering limited power capacity, uncertain inflexible loads, and PV generation. Here, we assume that EVs share information about their departure and the SoC is communicated at all times. The simulation uses the ``Workplace'' EV behavior model and has $85$, $15$-minute steps. Table~\ref{tab:comp_v2g_plus_loads} demonstrates an extensive comparison of all evaluation metrics after $100$ stochastic runs for every baseline algorithm. In this case study, profits and user satisfaction are the most important metrics. Here again, the optimal solution is not realistic but helps put an experimental upper bound. Notice that the total battery degradation of all EVs at the end of the simulations is measured for informational purposes, although it is not included in the objective or reward function of any model.

\begin{figure}[!t]
     \subfloat[Charging and discharging prices.]{
         \centering
         \includegraphics[width=0.97\linewidth]{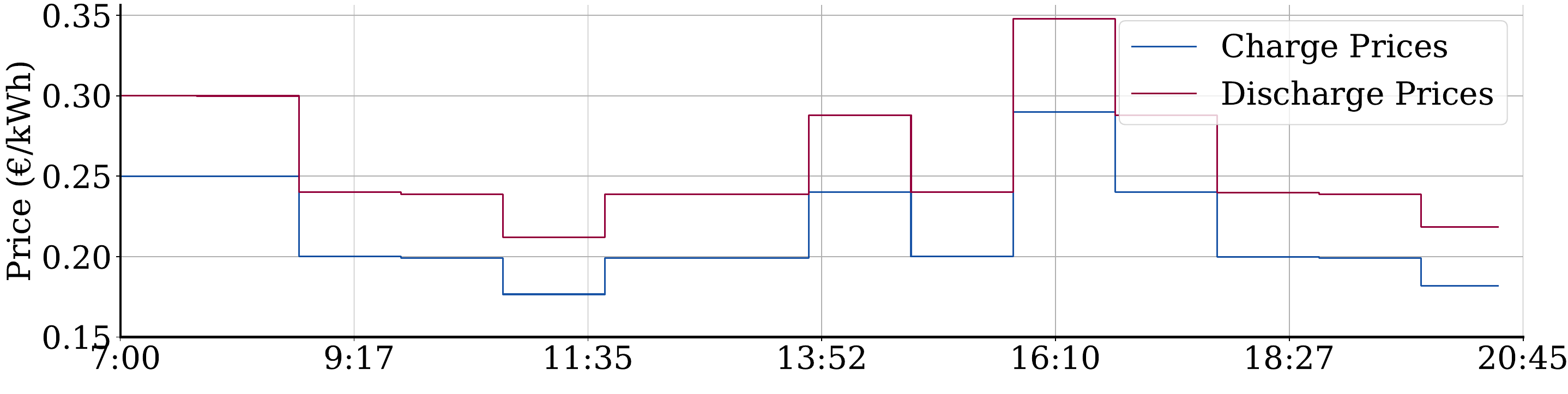}
         \label{fig:prices}
         }
         \\
     \subfloat[Total power comparison.]{
         \centering
         \includegraphics[width=0.97\linewidth]{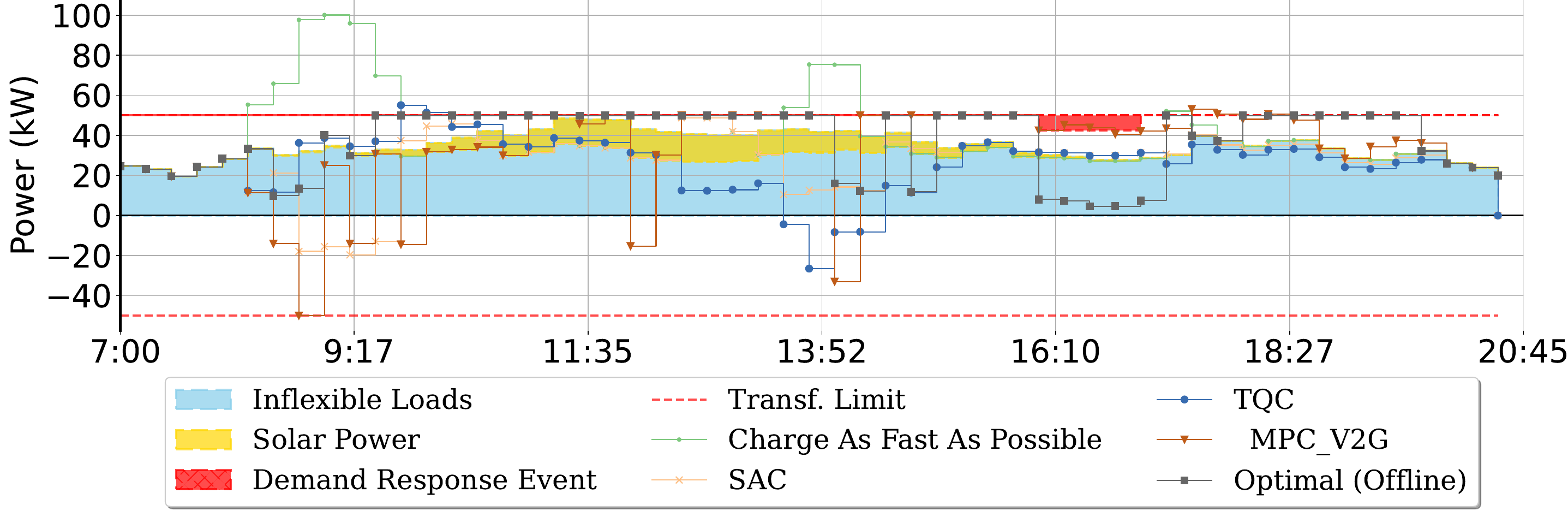}
         \label{v2gwithloads_comp}}
         \vspace{-2mm}
        \caption{ Performance comparison of selected algorithms in the V2G profit maximization with loads problem with $10$ bidirectional single-EVSE chargers.
        }
        \label{fig:v2gwithloads_comp_}
\end{figure}

As shown in Table~\ref{tab:comp_v2g_plus_loads}, no algorithm achieves $100\%$ user satisfaction while maximizing profits. AFAP, ALAP, and RR do not consider the energy prices and V2G; hence, they fail to maximize profits. MPC, utilizing a prediction and control horizon of $25$ steps, offers the most favorable balance between user satisfaction and profits. However, some RL methods, such as TD3, yield higher profits at the expense of reduced user satisfaction. Moreover, considering the execution time as a performance tradeoff is crucial. MPC requires significantly more time to determine optimal solutions, whereas RL algorithms are notably faster. The training time of RL is not considered in the comparison as it is done only once off-line. This underscores the limitations of mathematical programming solutions as problem complexity grows, thereby emphasizing the need for scalable solutions.

Fig.~\ref{fig:v2gwithloads_comp_} illustrates the performance of selected algorithms in comparison to the charging and discharging prices. Utilizing uncontrolled charging (AFAP) results in higher transformer overloads, whereas MPC and optimal solutions efficiently utilize the power transformer limits to their maximum capacity. Conversely, RL approaches appear to be more conservative in approaching the transformer power limits.

Overall, in both case studies, the results suggest that RL holds promise in offering effective solutions, albeit requiring some adjustments, such as modifying the reward or state functions, tuning algorithm hyperparameters, or exploring alternative algorithms. Nevertheless, EV2Gym stands ready to facilitate the development and comparative analysis of novel solutions, regardless of the approach chosen.

\begin{table*}[t]
\centering
\caption{Baselines comparison for the V2G profit maximization problem with $10$ EV chargers, transformer loads, PV, and DR events.}
\label{tab:comp_v2g_plus_loads}
\resizebox{\textwidth}{!}{%
\begin{tabular}{@{}ccccccccccc@{}}
\toprule
Algorithm & Profits/ Costs (\euro) & \begin{tabular}[c]{@{}c@{}} $\epsilon^{\textit{usr}}$ (\%)\end{tabular} & \begin{tabular}[c]{@{}c@{}}Energy \\Ch. (kWh)\end{tabular} & \begin{tabular}[c]{@{}c@{}}Energy \\ Disch. (kWh)\end{tabular} & \begin{tabular}[c]{@{}c@{}}Tr. Ov. \\ (kWh) \end{tabular} & \begin{tabular}[c]{@{}c@{}} Total $Q^\text{lost}$ \\ ($\times10^{-3}$)\end{tabular} &\begin{tabular}[c]{@{}c@{}} $\sum d^\cale$ \\ ($\times10^{-3}$)\end{tabular} &\begin{tabular}[c]{@{}c@{}} $\sum d^\cyc$ \\ ($\times10^{-3}$) \end{tabular}& \begin{tabular}[c]{@{}c@{}}Execution\\ Time (s)\end{tabular} & \begin{tabular}[c]{@{}c@{}}Reward\\ ($\times 10^{3}$)\end{tabular} \\ \midrule
AFAP & $-23.4$ ±$6.4$ & $100$ ±$0$ & $109$ ±$29$ & $0$ ±$0$ & $138$ ±$86$ & $0.40$ ±$0.09$ & $0.15$ ±$0.04$ & $0.24$ ±$0.07$ & $0.02$ ±$0.00$ & $-13.8$ ±$8.6$ \\
ALAP & $-25.2$ ±$7.1$ & $100$ ±$0$ & $109$ ±$29$ & $0$ ±$0$ & $106$ ±$78$ & $0.38$ ±$0.09$ & $0.14$ ±$0.04$ & $0.24$ ±$0.07$ & $0.02$ ±$0.00$ & $-10.6$ ±$7.8$ \\
RR & $-22.6$ ±$5.6$ & $99$ ±$1$ & $107$ ±$26$ & $0$ ±$0$ & $11$ ±$19$ & $0.45$ ±$0.11$ & $0.17$ ±$0.04$ & $0.29$ ±$0.08$ & $0.03$ ±$0.00$ & $-1.2$ ±$1.9$  \\ 
ARS & $17.7$ ±$9.9$ & $53$ ±$11$ & $45$ ±$19$ & $107$ ±$35$ & $36$ ±$67$ & $0.57$ ±$0.14$ &  $0.12$ ±$0.03$ & $0.45$ ±$0.11$ & $0.09$ ±$0.01$ & $-3.6$ ±$6.7$ \\
DDPG & $25.3$ ±$9.5$ & $44$ ±$9$ & $33$ ±$16$ & $125$ ±$34$ & $43$ ±$88$ & $0.59$ ±$0.13$ & $0.12$ ±$0.03$ &  $0.47$ ±$0.11$ & $0.12$ ±$0.02$ & $-4.4$ ±$8.8$ \\
PPO & $23.7$ ±$10.2$ & $46$ ±$9$ & $36$ ±$16$ & $123$ ±$38$ & $37$ ±$89$ & $0.58$ ±$0.14$ & $0.12$ ±$0.03$ &  $0.46$ ±$0.11$ & $0.13$ ±$0.27$ & $-3.7$ ±$8.9$ \\
TD3 & $31.1$ ±$10.6$ & $38$ ±$8$ & $25$ ±$13$ & $142$ ±$41$ & $74$ ±$140$ &  $0.61$ ±$0.13$ & $0.12$ ±$0.03$ &  $0.50$ ±$0.11$ & $0.12$ ±$0.03$ & $-7.5$ ±$14.0$ \\
TQC & $1.9$ ±$21.6$ & $72$ ±$24$ & $66$ ±$39$ & $62$ ±$58$ & $35$ ±$56$ & $0.53$ ±$0.13$ & $0.14$ ±$0.04$ &  $0.40$ ±$0.11$ & $0.16$ ±$0.02$ & $-3.5$ ±$5.6$ \\
TRPO & $13.5$ ±$8.5$ & $57$ ±$10$ & $41$ ±$17$ & $88$ ±$30$ & $7$ ±$13$ &  $0.52$ ±$0.12$ & $0.14$ ±$0.03$ & $0.40$ ±$0.09$ & $0.13$ ±$0.01$ & $-0.7$ ±$1.3$ \\
MPC & $17.8$ ±$9.3$ & $76$ ±$9$ & $480$ ±$107$ & $462$ ±$109$ & $183$ ±$167$ & $1.68$ ±$0.42$  & $0.12$ ±$0.03$ &  $1.55$ ±$0.39$ & $108.18$ ±$19.44$ & $-18.3$ ±$16.7$ \\ \midrule
Optimal & $3.7$ ±$6.1$ & $100$ ±$0$ & $578$ ±$135$ & $469$ ±$117$ & $4$ ±$12$ & $1.45$ ±$0.35$  & $0.15$ ±$0.04$ & $1.31$ ±$0.32$ & $33.30$ ±$28.05$ & $-0.4$ ±$1.2$ \\ \bottomrule
\end{tabular}%
}
\end{table*}



\section{Conclusions}
\label{sec:sec6}

In this paper, we introduced EV2Gym, an innovative V2G simulator designed to address critical gaps in the development and evaluation of smart EV charging algorithms. Unlike existing tools, EV2Gym integrates realistic CPO assumptions and detailed models of EVs, charging stations, and EV behaviors, leveraging real open-source data to enhance fidelity.
Finally, we demonstrated the user-friendly process of designing new case studies or custom algorithms and leveraging existing baseline algorithms to accelerate the evaluation process.


\bibliographystyle{IEEEtran}
\bibliography{ref}

\end{document}